\pdfoutput=1
\documentclass[fleqn, 12pt]{scrreprt}
\usepackage[english]{babel}
\usepackage[latin1]{inputenc}

\usepackage{amsmath, amsfonts, amssymb, amsthm} 
\usepackage{mathrsfs} 
\usepackage{xfrac}  
\usepackage{cases}  

\usepackage{abstract}  
\usepackage{appendix}  
\usepackage{tocvsec2}  
\usepackage{booktabs}  
\usepackage{rotating}  
\usepackage{pbox}      
\usepackage{chngcntr}  
\counterwithout{footnote}{chapter}

\usepackage{listings} 
\usepackage[capposition=bottom]{floatrow} 
\usepackage{placeins}  

\usepackage{blindtext} 
\usepackage{xspace}    
\usepackage{setspace}  

\usepackage{hyperref} 
\usepackage{cleveref} 

\author{Lars Roemheld}
\date{Wintersemester 2012/13}

\subject{Bachelor Thesis}
\title{Evolutionary Extortion and Mischief}
\subtitle{Zero determinant strategies in iterated 2x2 games}
\publishers{Supervised by Prof. Dr. Joerg Oechssler}

\begin{document}
  \pagenumbering{gobble} 
  
  \begin{titlepage}
    \begin{center}
 
    \Large\textbf{Heidelberg University \\ Alfred-Weber-Institute for Economics}
    \vspace{5cm}

    \normalsize
    Bachelor Thesis\\
    
    \LARGE\textbf{Evolutionary Extortion and Mischief}\\
    \Large\textbf{Zero Determinant strategies in iterated 2x2 games}\\
    
    \vspace{3.0cm}
    submitted by\\
    \vspace{0.5cm}
    \Large{Lars Roemheld}\\
    \normalsize
    born in Duisburg (Germany)
    
    \vspace{2cm}
    \normalsize
    Supervised by: Prof. Dr. Joerg Oechssler
    
    \vspace{5.0cm}
    \Large\textbf{January 2013}

    \end{center}
  \end{titlepage}

	\newcommand{\aUP}{\text{\texttt{up}} \xspace}
	\newcommand{\aDOWN}{\text{\texttt{down}} \xspace}
	\newcommand{\aUU}{\text{uu} \xspace}
	\newcommand{\aUD}{\text{ud} \xspace}
	\newcommand{\aDU}{\text{du} \xspace}
	\newcommand{\aDD}{\text{dd} \xspace}
	\newcommand{\aClockWise}{\aUU, \aUD, \aDU, \aDD}

	\newcommand{\aAllU}{\text{AllU} \xspace}
	\newcommand{\aAllD}{\text{AllD} \xspace}
	\newcommand{\aTFT}{\text{TFT} \xspace}

	\newcommand{\mem}[1]{\mbox{memory-#1}}
	\newcommand{\PrDy}{Press and Dyson\xspace}
  
	\newcommand{\mymarker}{\rule{2cm}{1em}}

	\renewcommand{\vec}[1]{\pmb{#1}}
	\newcommand{\mat}[1]{\textbf{#1}}
	\newcommand{\T}{\intercal}
  
	\newcommand{\mydots}{\:\dotsc\:}
  
  \renewcommand{\labelitemi}{\raisebox{.45ex}{\rule{.6ex}{.6ex}}}
  \renewcommand{\emph}[1]{\textsc{#1}}

	\begin{abstract}
    \noindent This paper studies the mechanisms, implications, and potential applications of the recently discovered class of Zero Determinant (ZD) strategies in iterated 2x2 games. These strategies were reported to successfully extort pure economic maximizers, and to mischievously determine the set of feasible long-term payoffs in iterated Prisoners' Dilemma by enforcing linear constraints on both players' expected average scores.
    \vspace{5pt}
    
    These results are generalized for all symmetric 2x2 games and a general Battle of the Sexes, exemplified by four common games. Additionally, a comparison to conventional strategies is made and typical ZD gameplay simulations are analyzed along with convergence speeds. Several response strategies are discussed, including a glance on how time preferences change previous results. Furthermore, a possibility of retaliation is presented: when maximin scores exceed the minimum symmetric payoff, it is possible to extort the extortioner.
    \vspace{5pt}
    
    Finally, a summary of findings from evolutionary game theory shows that mischief is limited by its own malice. Nevertheless, this does not challenge the result that mindless economic maximization is subject to extortion: the study of ZD strategies reveals exciting new perspectives and opportunities in game theory, both evolutionary and classic.

    \vfill
	\end{abstract}

	\pagenumbering{roman} 
  \tableofcontents

  \parskip3pt
  \setstretch{1.3}  
  \setlength{\parindent}{0.5cm}  

  \widowpenalty=10000
  \clubpenalty=10000

	\chapter{Introduction}
  \label{chap:Intro}
	\pagenumbering{arabic} 

In many branches of the social sciences, game theory has been used to model, predict, and understand interaction of a number of agents subjected to the same rules: games can be anything from auctions to competition, from political elections to social contracts, and from judiciary to matrimonial trouble. Perhaps the simplest useful class of games are 2x2 games: interactions between two players, each choosing one of two actions. Despite their simple appearance, many phenomena of game theory can already be observed in this basic form: ``the players act rationally, but the consequences are often bizarre, which makes application to a world of intelligent men and ludicrous outcomes appropriate'' \cite[p. 2]{RasmusenGT}.

The contrast between the same 2x2 games when they are played once and when they are played in series, both in simplicity of solution concepts and in their strategic implications, led to a wide range of social phenomena being rationalized: the same egoistic players who exhibit rational yet (bizarrely) anti-social behavior in once-off encounters can be driven to more cooperative actions in iterated games. Consequently, the theory of 2x2 games has often been used as a study of social cooperation \cite{AxelrodCoop}.

In a recent paper by William Press and Freeman Dyson \cite{PrDy}, the class of Zero Determinant (ZD) strategies for iterated 2x2 games was discovered, promising to beat rationalistic agents on their own ground: mindful players can exploit pure economic rationality by extorting a larger payoff share for themselves. The aim of this present paper is to study those strategies, which were believed to threaten the more cooperative nuances in game theory. How, and under what conditions do ZD strategies work? Can they indeed outperform the intuitive and conventional behavior of tit for tat, as has been claimed? How can we gain from this new and mischievous knowledge? And how can the less mischievous protect themselves? 

This chapter provides a minimal context to understand the foundations ZD strategies stand on. All analysis in this paper strives to be as general as possible; however, to facilitate demonstration and to provide meaningful examples, the first section will introduce four of the most common 2x2 games, noting some of their applications. When such stage games are iterated, their dynamic changes substantially: the following sections thus introduce the theory of iterated 2x2 games, and one of its landmarks, the Folk theorem. Having seen that predictions about iterated games are less clear-cut than their once-off pendants, a short summary of simulated tournaments between different strategies in iterated games will complete this chapter.

In \cref{chap:ZDStrat}, I analyze the mechanisms and implications of ZD strategies. After a thorough recapitulation of their derivation, they be will generalized for more general games and their behavior will be studied both analytically and numerically. Finally, I will provide a number of response options and try to integrate Zero Determinant strategies with classic game theory.

In \cref{chap:EvolComp} these results are utilized for a brief digression to evolutionary game theory, discussing potential applications of ZD strategies as well as their chances of success.

Finally, a short commentary will conclude this paper. In the appendices, I provide technical details for my results.

	\section{Exemplary 2x2 stage games}
  \label{IntroStageGames}

Four of the most commonly studied 2x2 games will serve as examples in this paper: the Prisoners' Dilemma, Stag Hunt, the Game of Chicken, and the Battle of the Sexes. The first three of those are symmetric games, which means that they can all be represented in a symmetric payoff matrix (see \cref{tab:symmetricGame}): this reflects the fact that players have the same preferences over game outcomes, as defined by their own and their opponent's move.

\begin{table}[hbtp]
	\centering
  \begin{tabular}{l|c|c|}
    \multicolumn{1}{l}{}  &  \multicolumn{1}{c}{$\aUP$} 	& \multicolumn{1}{c}{$\aDOWN$} \\
    
    \cline{2-3}
    $\aUP$ & R, R & S, T \\
    \cline{2-3}
    $\aDOWN$ & T, S & P, P \\
    \cline{2-3}
  \end{tabular}
	\caption{A symmetric 2x2 game payoff matrix}
	\label{tab:symmetricGame}
\end{table} 
In any 2x2 stage game, two players face a once-off decision between two possible actions: in this example they are generically named ``$\aUP$'' and ``$\aDOWN$''. One player, $X$, chooses an action on the table's rows (hence she is called row-player), while the other, $Y$, chooses an action on the table's columns (hence he is called column-player). If the row-player chooses her action $a_X=\aDOWN$ and the column-player chooses his action $a_Y=\aUP$, for instance, the outcome of the game shall be referred to as $(a_X, a_Y)=(\aDOWN, \aUP)$. Payoffs for player $i \in \{X, Y\}$ are formally defined as a function giving a score for each possible outcome: $\pi_i : (a_X, a_Y) \rightarrow \mathbb{R}$. In the example of the symmetric game, payoffs for $(\aDOWN, \aUP)$ are then $\left(\pi_X(\aDOWN, \aUP), \pi_Y(\aDOWN, \aUP)\right) = (T, S)$: the cells in \ref{tab:symmetricGame} show payoffs $(\pi_X, \pi_Y)$ for each of the possible game outcomes. Players are assumed to have strongly monotone preferences over payoffs, implying $T>S \Leftrightarrow T \succ S$ where ``$\succ$'' is the strong preference relation.

I assume quantitative symmetry in this paper, meaning that symmetry not only holds for ordinal relations between payoffs but rather for the exact payoff values. That is, $\pi_X(a_X, a_Y) = \pi_Y(a_Y, a_X)$. Here, this allows for easier notation without loss of analytic generality.

Finally, for completeness, we loosely define:%
\footnote{%
          For more detailed definitions, see e.g\@. \cite{GibbonsGT}.
          }
a \textit{strategy} $\sigma_i$ of player $i$ is a ``plan to choose an action $a_i$''. The payoff function can then be defined over strategies instead of actions: $\pi_i : (\sigma_X, \sigma_Y) \rightarrow \mathbb{R}$. A \textit{Nash equilibrium} is a combination of strategies for both players where, given the strategy of the other player, no player would want to change his strategy. A \textit{maximin strategy} is a strategy for player $i$ yielding his \textit{maximin payoff} $r := r_i = \max_{\sigma_i} \min_{\sigma_{-i}} \pi_i(\sigma_X, \sigma_{Y})$ (where $-i$ signifies player $i$'s opponent): the highest payoff player $i$ can guarantee himself regardless of his opponent's action.%
\footnote{%
          Since we analyze quantitatively symmetric games, both players have the same maximin payoff. To simplify notation, I use $r=r_X=r_Y$.
          }
In this paper, maximin is always ``pure strategy maximin,'' i.e\@. restricted to strategies that explicitly define an action $a_i$. As for notation: $\vec{v}$ is a column vector, $\vec{v}^\T$ is the transposed vector (a row vector), $v_k$ is the $k$-th component of $\vec{v}$, $\mat{N}$ is a matrix, $p(A)$ is the probability of $A$, $p(A|B)$ is the conditional probability of $A$, conditioned on $B$, and $E[A]$ is the expected value of $A$.

	\subsection{Prisoners' Dilemma}
	
The Prisoners' Dilemma (PD) is probably the most widely studied game in game theory; it is characterized by the relations $T>R>P>S$ and $2R>T+S$. In this paper, Axelrod's canonical payoffs \cite{AxelrodCoop} will be used for numerical analyses, i.e\@. $(R,S,T,P)_{PD}=(3,0,5,1)$.

While its original story is built around a federal leniency program, PD is generally treated as a game about cooperation: suppose two parties share a common good. Both parties can choose to either respectfully use that good in moderation (cooperate, $\aUP$), allowing both parties to benefit from it. Or they can claim an unproportional share (defect, $\aDOWN$), which promises higher returns regardless of their opponent's action. Due to the individual incentive to defect, both parties find themselves in mutual defection, resulting in the shared good being destroyed. The mechanism of individual incentives favoring a collectively disappointing outcome can be applied to a wide range of phenomena, from free riding to the tragedy of the commons, from the rule of law to pollution. Because of its universal character and its intuitive simplicity, PD's general structure has become paradigmatic to the study of social cooperation: indeed, even contemporary social philosophers continue to rely on its logic \cite[pp. 238ff.]{Rawls}.

Because of the incentive to defect regardless of the opponent's action, PD has a single Nash equilibrium, $(\aDOWN, \aDOWN)$, which is usually called ``mutual defection.'' The resulting payoffs are also the maximin payoff: $r=P$.

	\subsection{Stag Hunt}
	
Stag Hunt (SH) is another game often referred to in the context of cooperation. In SH, the following relation holds: $R>T\geq P>S$. To make the game more interesting for illustrative purposes, I assume $T=P$. Then two Nash equilibria exist in pure strategies: $(\aUP, \aUP)$ and $(\aDOWN, \aDOWN)$, where the former is payoff dominant (players receive higher payoffs) and the latter is risk dominant (players do not risk being exploited, for they receive the same payoff regardless of their opponent's action). Obviously, the risk dominant Nash equilibrium also gives the maximin payoffs, $r=P$. Additionally, a mixed Nash equilibrium exists, where both players choose to play $\aUP$ with a certain probability.

When numerical analyses are required, I will use $(R, S, T, P)_{SH}=(10, 0, 8, 8)$. With these payoffs, the mixed Nash equilibrium is for both players to choose $p(\aUP)=0.8$---the players are then indifferent between both pure actions.

	\subsection{Game of Chicken}
	
The Game of Chicken (GC) has been used to study escalating behavior in conflicts and has famously been applied to brinkmanship in nuclear warfare \cite[pp. 30ff.]{Russell}. In GC, payoffs are characterized by $T>R>S \geq P$. For numerical values, I take $(R, S, T, P)_{GC}=(6, 2, 7, 0)$.

In GC, both players ``dare each other'' to play $\aDOWN$, but prefer any alternative to $(\aDOWN, \aDOWN)$ (the worst result where both players ``escalate''). Thus two pure strategy Nash equilibria exist, $(\aUP, \aDOWN)$ and $(\aDOWN, \aUP)$---one player will always prefer to back down. Additionally, a mixed Nash equilibrium exists (with payoffs as defined, both players choose $p(\aUP)=2/3$). In GC, the maximin payoff is $r=S$, the minimum payoff from backing down.

	\subsection{Battle of the Sexes}

As the most common example of asymmetric games, this paper further studies the Battle of the Sexes (BS): here the players have different preferences about the game's outcome, yielding a different payoff matrix (see \cref{tab:BoSGame}). Payoff values satisfy $F > D > C \geq L$, and I will assume $(F, C, L, D)_{BS}=(5, 1, 1, 3)$ for numerical analyses.

\begin{table}[h]
	\centering
  \begin{tabular}{l|c|c|}
    \multicolumn{1}{l}{}  &  \multicolumn{1}{c}{$\aUP$} 	& \multicolumn{1}{c}{$\aDOWN$} \\
    
    \cline{2-3}
    $\aUP$ & F, D & C, C \\
    \cline{2-3}
    $\aDOWN$ & L, L & D, F \\
    \cline{2-3}
  \end{tabular}
	\caption{Payoffs in the Battle of the Sexes}
	\label{tab:BoSGame}
\end{table}
The dilemma of a couple who cannot agree on a way to spend the evening%
\footnote{%
          ``While selfish, they are deeply in love and would, if necessary, sacrifice their [favorite alternative] to be with each other. Less romantically, their payoffs are [...]'' \cite[p. 28]{RasmusenGT}
          }
epitomizes coordination issues between two parties: imagine negotiations over contractual conditions: both parties want the contract to be signed, but both hope to establish their preferred alternatives. Pure strategy Nash equilibria thus are $(\aUP, \aUP)$ and $(\aDOWN, \aDOWN)$ (successful coordination), and a mixed Nash equilibrium exists (with payoffs as defined, both players choose $p(\aUP)=1/3$). The maximin payoff is $r=C$, i.e\@. failing to coordinate but having claimed the preferred alternative.

	\section{Iterated games}
	\label{sec:itGames}

Many applications of 2x2 stage games are better described when the assumption of once-off encounters is dropped: instead, two players repeatedly play the same game against each other, each time receiving a stage game payoff. 

Suppose the Prisoners' Dilemma is played repeatedly. Then $PD(T)$ signifies the iterated game obtained by playing PD $T$-times. Let $PD_t$ signify the $t$-th period of $PD(T)$, i.e\@. the $t$-th sequential stage game. Each stage game $PD_t$ is the same, except for a history of game outcomes that were already played, and a different number of iterations, $PD_{t'>t}$, to follow. Both players know this, i.e\@. they know how many more ``rounds'' they will play, and they recall a history of previous outcomes. Let $H_{t}:=\left( (a_X, a_Y)_1, \mydots , (a_X, a_Y)_t \right) $ be such a history of $t$ outcomes, signified by the actions $a_X, a_Y \in \{\aUP,\aDOWN \}$ in $PD_{1, \mydots , t}$. This shared knowledge allows both players to condition their action in $PD_t$ on the observed history of previous outcomes: their strategies can be understood as functions of history, $\sigma_i(H_t)$.

If both players do not know exactly when the sequence of stage game iterations is going to end (i.e\@. there is no clear-cut $T$, or the players don't know its value), the iterated game may be analyzed as an infinitely iterated game, $PD(\infty, \delta)$, with a certain probability%
\footnote{%
          Without loss of generality, let this also cover players' time preferences: no additional discount rate is considered in this paper. In the following analyses, $G(\infty, \delta)$ is to mean ``infinitely repeated stage game $G$ with common discount rate $\delta$.''}
$(1-\delta)$ of ending in every period \cite[pp. 131ff.]{RasmusenGT}. Under such circumstances, no player can ever be certain that he is playing the last round of the iterated game. Aside from the history of previous plays, every iteration thus presents the same stage game: the probability of the game continuing is always the same. Analysis shows that the clear-cut Nash equilibria of stage games are now complemented by a multitude of additional equilibria: this will be demonstrated for the infinitely iterated PD (iPD---iSH, iGC, and iBS conversely are the infinitely iterated counterparts of the stage games defined above) before the more general Folk theorem is introduced in the next subsection.

The once-off PD game, where ``mutual defection'' is the only Nash equilibrium, is the archetype of rational players and bizarre outcomes. By contrast, the ``cooperative'' game outcome $(\aUP, \aUP)$ can be stabilized as Nash equilibrium for \emph{every} iteration of iPD \cite[pp. 216ff.]{EichbergerGT}; consider player $Y$ in one iteration of iPD. Knowing that he will probably continue to play a sequence of games after this one, he will want not only to myopically maximize his payoff in this iteration, but also to consider future payoffs. His interest in future payoffs, to be precise, will depend on the discount factor $\delta$, as he will try to maximize the discounted payoff $\mathscr{P}_Y$, where 

\begin{equation}
  \mathscr{P}_i (\sigma_X, \sigma_{Y}) := \liminf_{T \rightarrow \infty} \sum^{T}_{t=1} \delta^{t-1} \pi_i(\sigma_X(H_t), \sigma_{Y}(H_t))
  \label{eq:DiscountedPayoff}
\end{equation}
The discount factor simply reduces the attention paid to payoffs from more distant (and therefore less likely) iterations.

Now assume $X$ to play the following grim trigger strategy: play $\aUP$ until $Y$ plays $\aDOWN$ for the first time. Then forever continue playing $\aDOWN$. In every iteration $Y$ can choose to either receive $T$ followed by an infinite series of $P$, or to receive an infinite series of $R$. If $Y$ cares enough about future payoffs (i.e\@. if $\delta$ is large enough), the future cost of reverting to the stage game equilibrium will outweigh the present benefit of defecting, and playing $\aUP$ in every iteration will maximize $Y$'s overall score. Then if $Y$ decides to play the analogous grim trigger strategy ($\aUP$ until $X$ first plays $\aDOWN$), neither player has an incentive to deviate from their strategies, yielding $(\aUP, \aUP)$ in every iteration of iPD.

This result has been understood as key to social phenomena: even in situations which normally discourage cooperation, such as the tragedy of the commons \cite{DawesDilemmas} as exemplified in iPD, cooperation becomes possible if two players do not know how long they will interact with each other, and if they care enough about future interactions.

	\section{The Folk theorem}
  \label{sec:Folk}
  \FloatBarrier
	
The Folk theorem generalizes the result that by iterating a stage game infinitely, more Nash equilibria are obtained than in the stage game. To be precise, applied to 2x2 games, it states that in any infinitely iterated game $G(\infty, \delta)$ there exist (subgame-perfect \cite[pp. 221f.]{EichbergerGT}) Nash equilibria that yield average stage game payoffs of $(\pi_X, \pi_Y)$ if the following conditions hold \cite[p. 97]{GibbonsGT}:

\begin{enumerate}
	\item {The stage game $G$ has a Nash equilibrium yielding payoffs $(e_X, e_Y)$, and $\pi_i>e_i$ for both players $i$.}
	\item {The probability of the game continuing (the discount rate), $\delta$, is sufficiently close to 1. This ensures that threats about future behavior have a big enough impact on present decision making.}
\end{enumerate}
Analogous to the PD example above, in any infinitely iterated game a credible threat can be made to fall back to the Nash equilibrium yielding payoffs $(e_X, e_Y) \ll (\pi_X, \pi_Y)$, a worse (but equilibrium) outcome for both players. A grim trigger strategy would thus be along the lines of ``play a combination of stage game actions yielding on average $(\pi_X, \pi_Y)$, until an opponent diverts from that combination. Then, continue playing the stage game equilibrium strategy yielding $(e_X, e_Y)$ forever after.''

\begin{figure}[hbt]
  \centering
  \def\svgwidth{10cm}
  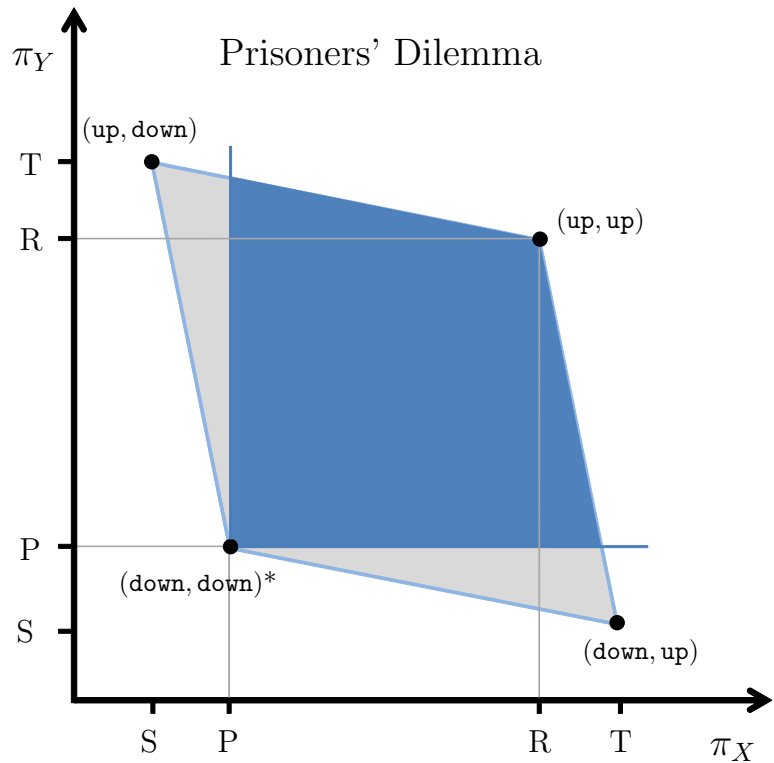
  \caption{The Folk theorem: feasible Nash equilibria in iPD}
	\label{fig:FolkConvexSet}
  \floatfoot{Note: the set of possible payoffs is the convex hull of all stage game payoffs, the total gray and blue area. The Folk theorem states that the area marked blue can be achieved as subgame-perfect Nash equilibria in the infinitely iterated game by threatening to play the stage game Nash equilibrium (*).}
\end{figure}

This result adds a convex set of feasible Nash equilibria to the infinitely iterated game: \cref{fig:FolkConvexSet} demonstrates this in the case of iPD. All strategy combinations yielding payoffs in the blue area can be achieved as equilibrium payoffs in iPD when $\delta$ is large enough. However, the Folk theorem merely states that all combinations in this set \emph{can} be achieved as equilibria; no statement is made about whether such equilibria \emph{will} indeed be reached.

	\section{Tournaments of strategy zoos}
  \label{sec:Tournaments}

Infinitely iterated 2x2 games have been used to study a wide range of topics, from social behavior (e.g\@. \cite[pp. 73ff.]{AxelrodCoop}) to economic competition (e.g\@. \cite[pp. 136ff.]{RasmusenGT}) and even questions of evolutionary biology (see e.g\@. \cite{Weibull} for the diverse literature). The Folk theorem gives some indication on why their relatively simple mechanism (two players, each having only two possible actions) has sparked so much interest: while it is clear that a whole set of payoffs can generally be achieved as Nash equilibrium, it remains unclear which specific point from the set will ensue in a game and which strategies are suited to obtain optimal payoffs when no information is available about one's opponent.

The question therefore is which strategies would be most successful if infinitely iterated games were played ``in the wild.'' In diverse applications, formal analyses as well as computer simulations are applied to find especially successful strategies, or to understand the dynamics of different populations of strategies (representing for example cultural conventions, marketing strategies, or biological species). 

In computer simulations, populations of players using different strategies are randomly pitted against each other to form a competitive, unpredictable environment. The most famous of such tournaments, Axelrod's first iPD tournament \cite{AxelrodCoop}, showed variations of the so called tit-for-tat strategy ($\aTFT$) to be particularly successful. The original $\aTFT$ strategy plays $\aUP$ in the first iteration and then plays whatever move its opponent played last; much like the grim trigger strategy, $\aTFT$ can successfully establish cooperative payoffs. 

In subsequent tournaments, both amateur \cite{LessWrong} and scientific, this result was confirmed, consistently ranking more or less ``cooperative'' strategies similar to $\aTFT$ or win-stay-loose-shift \cite{WinStay} among the top strategies in terms of total payoffs: while they can generally be exploited by an opponent primed to do so, they tend to achieve relatively high payoffs both for themselves and for their opponents. This makes them typical winners in tournaments of strategy zoos when it comes to total payoffs earned, while ``losing'' most individual games against single opponents: while more ``fierce'' strategies can outperform $\aTFT$ by a small amount, they will receive much smaller payoffs in the fierce battle between themselves, enabling $\aTFT$ to win by total payoff earned.

These results are confirmed in a wide range of literature (for an overview, see \cite{AxelrodComp,Hoffmann}), and for strategies like $\aTFT$ it has been established that they can only be beaten by an infinitesimal amount, effectively equalizing their own and their opponent's score.

	\chapter{Zero Determinant Strategies}
  \label{chap:ZDStrat}

The well-established success of such intuitive patterns as tit-for-tat and the ``fairness'' they induce contributed to the considerable attention received by a paper recently published by William Press and Freeman Dyson \cite{PrDy}:%
\footnote{%
          Besides a number of scientific responses published so far, the paper was notably covered in popular media, despite its technicality \cite{ZEITreport,Chronicle,TechnologyReview}.
          }
they report the discovery of a class of ``simple ultimatum [strategies] whereby one player can enforce a unilateral claim to an unfair share of rewards'' \cite[p. 1]{PrDy} in iPD. The properties of their Zero Determinant (ZD) strategies are this paper's main interest.

In \cref{sec:Generality,sec:gamesAsMarkov}, we follow \PrDy's derivation of ZD strategies, expanding on critical points and introducing the notation used in this chapter. \Cref{sec:MischiefZD} interprets \PrDy's main results, which are generalized and illustrated in \cref{sec:ZDparameters}. There, concrete strategies are shown for the four games analyzed here, along with results from exemplary gameplay simulations.

The implications for a ZD player's opponent are discussed in \cref{sec:BestResponse}, where I present conditions for a simple yet effective response to ZD strategies. To assess the aptitude of ZD strategies in real-world applications, \cref{sec:ZDConvergence} analyzes convergence speed. Finally \cref{sec:ZDFolk} attempts to integrate ZD strategies with classical game theoretic approaches.

	\section{Generality of \mem{1} analysis}
  \label{sec:Generality}

In \cref{sec:itGames} it was assumed that players in an iterated game know the whole history of plays; however, especially in infinitely iterated games, it appears sensible to assume that players condition their strategy only on a finite subset of recent history---if this is the case, competing players might use different memory spans: one player who remembers and uses the last 20 outcomes may compete against a particularly forgetful player who can only recall the very last move, and thus only conditions his strategy on the very last outcome. The strategy that is conditional on the last 20 moves will be called a \mem{20} strategy; the latter players' strategy is a \mem{1} strategy. Note that even a player with long memory might play a \mem{1} strategy, i.e\@. she might condition her moves on the last outcome only.

ZD strategies are a subclass of \mem{1} strategies, i.e\@. they are conditional only on the last stage game's outcome. Since in their analysis \PrDy pit them against a general \mem{1} strategy, they have to show that this is sufficiently general to cover \emph{all} possible opponents, including those who have longer memory. They do this by showing---somewhat counterintuitively---that the ``shortest memory-player sets the rules of the game'' \cite[p. 4]{PrDy}: playing any \mem{$n$} strategy would not yield any advantages over playing another \mem{1} strategy against ZD strategies. This refutes the obvious argument against their analysis, viz\@. that it fails to cover more intricate opponents.

Any memory in excess of the shared memory of both players is irrelevant. This is due to the simple fact that a \mem{1} strategy player cannot apprehend reactions conditioned on a history she forgot already. Thus, to her, even the most intricate strategy will appear to be a \mem{1} strategy: the one obtained by averaging over all outcomes remembered by her opponent but not by her. If her opponent were to play this averaged-over strategy, she would not notice any difference to the \mem{$n$} strategy. Since her strategy will thus not be affected by averaging, the relevant gameplay would be equivalent: the distribution over the number of stage game outcomes, 
\begin{equation}
\sum_{g \in \mathbb{S}}{\ \sum_{C_g=0}^{\infty} {p(C_g)}}=1
\end{equation}
where $C_g$ is the number of stage game outcomes $g \in \mathbb{S}$ (such as $(\aUP, \aDOWN)$, see below) in the infinitely iterated game, will be exactly the same.

\PrDy present an analytical proof \cite[appx 1]{PrDy}; this somewhat counter-intuitive result can further be made plausible by two arguments. Firstly, memory enables strategy patterns, such as ``play $\aDOWN$ if the opponent played $\aDOWN$ twice in a row.'' If, however, one player's strategy pattern is a sequence longer than his opponent's memory, the pattern length in excess of his opponent's memory will never be noticed by his opponent, and will thus never influence his actions. Secondly, an intuitive objection is that the longer-memory opponent will not know the outcomes to average over before the game begins. While this is true, it is besides the point: after the game, we know that a shorter-memory strategy existed that would have yielded the same gameplay---during the game, the players just did not necessarily know which one. 

Thus, if a \mem{1} strategy is analyzed against another \mem{1} strategy, all possible game outcomes will be considered: all possible strategies that could ever play against a ZD strategy \textit{ex post} have a corresponding \mem{1} strategy that yields the same payoffs. Therefore, \PrDy's analysis of iPD as a game between two \mem{1} strategies is sufficiently general to cover all possible outcomes in iPD: opponents with longer memory cannot ``outplay'' ZD strategies.

	\section{Infinitely iterated games as Markov processes}
  \label{sec:gamesAsMarkov}

Having established the generality of \mem{1} analysis, this section follows \PrDy in their derivation of ZD strategies, expanding on theoretical background. Along with \PrDy, we can without loss of generality describe any infinitely iterated 2x2 game with at least one ZD strategy player by the following:%
\footnote{%
          For a mathematically more rigorous account of the derivation of ZD strategies, see \cite{Akin}.
          }

\begin{itemize}
	\item {Two \mem{1} players, $X$ and $Y$.}
	\item {
		Both players' action sets and the resulting set of possible stage game outcomes as described by $(a_X, a_Y) \equiv\ \text{xy} \in ( \aClockWise ) =: \mathbb{S}$, where we have defined $\aUD:=(\aUP, \aDOWN)$, and $\aUU, \aDU, \aDD$  analogously for notational convenience.
	}
	\item {
		Each player's strategy, which is defined as a vector of conditional probabilities of playing $\aUP$, conditioned on the last stage game outcome (the ``content'' of each player's \mem{1}). Since both players decide between $\aUP$ and $\aDOWN$ only, the probability of playing $\aDOWN$ is given by 
    $p(a_i=\aDOWN | \text{xy}) = 1 - p(a_i=\aUP | \text{xy})$.
		
		$X$'s strategy is $\vec{p}=(p_1,p_2,p_3,p_4)^\T$ where the order of probabilities corresponds to the last game outcome, $\text{xy} \in (\aClockWise)$ (i.e\@. $p_2:=P(a_X=\aUP | \text{xy} = \aUD)$). Likewise, $Y$'s strategy is $\vec{q}=(q_1,q_2,q_3,q_4)^\T$ where the probabilities correspond to the last game outcome, seen from his perspective, $\text{yx} \in (\aClockWise)$. Thus, $q_2:=P(a_Y=\aUP | \text{xy}=\aDU)$. Then the simplest strategies are the unconditional $\aUP$-player, $\aAllU$ with $\vec{p}=(1, 1, 1, 1)^\T$, and the unconditional $\aDOWN$-player, $\aAllD$ with $\vec{p}=(0, 0, 0, 0)^\T$
		
		Technically, to be complete, both strategies need a fifth, unconditional probability of playing $\aUP$ in the first move (when no history is known to condition the move on). However, since the game is infinitely long, the first move is of neglectable relevance for \PrDy's analysis \cite{SigmundNowakExtort,CalculusSelfishness}. This will especially hold true in the following analysis using Markov chain theory.
	}
	
	\item {Payoff vectors for $X$ and $Y$, in order of $ \text{xy} \in (\aClockWise)$. In the example of symmetric games, $\vec{s_X}=(R, S, T, P)^\T$ and $\vec{s_Y}=(R, T, S, P)^\T$ (see \cref{tab:symmetricGame}).
	}
  
  \item {
    A common discount factor for both players. To facilitate analysis, and in line with \PrDy's (implicit) practice, we assume $\delta=1$ for now. This allows us to assume that both players seek to maximize their (expected) average payoff over all iterations, which is easily comparable to simple stage game payoffs, since an infinite series of stage games yielding payoffs $R$ will have an average payoff of exactly $R$. Maximizing average payoff is equivalent to maximizing total payoff (which is given by \cref{eq:DiscountedPayoff} for $\delta=1$) \cite[pp. 210ff.]{EichbergerGT}.
    
    Since the average payoffs are directly comparable to stage game payoffs, we will later define $\pi_i$ to denote player $i$'s expected average payoffs, continuing the same notation from \cref{chap:Intro}: from here on, only infinitely iterated games will be analyzed, and $\pi_i$ is simply taken to mean ``expected average score over all, i.e\@. infinitely many games to come.'' The (strong) assumption of $\delta=1$ will be loosened later (see \cref{sec:BestResponse}).
  }
\end{itemize}
An infinitely iterated game thus defined is equivalent to a time-homogenous%
\footnote{%
          This assumes that for each player an equivalent, fixed \mem{1} strategy exists (i.e\@. there are equivalent $\vec{p}, \vec{q}$ that remain constant over all iterations). This can be shown to be true (cf. \cref{sec:ZDConvergence})}
Markov-chain over four possible states, $\mathbb{S}=(\aClockWise)$. This means that the sequence of stage game outcomes can be analyzed as a stochastic process where the probability of the next state depends only on the current state. If this game's outcome was $(ud)$, for instance, the probability of the next game's outcome being $(du)$ is $P(du | ud)=P(a_X=\aDOWN|ud)*P(a_Y=\aUP|ud) \equiv (1-p_2)q_3$. This yields the iterated game's Markov transition matrix:%

\begin{equation}
	\mat{M}=
	\begin{bmatrix}
		p_1q_1 & p_1(1-q_1) & (1-p_1)q_1 & (1-p_1)(1-q_1)\\
		p_2q_3 & p_2(1-q_3) & (1-p_2)q_3 & (1-p_2)(1-q_3)\\
		p_3q_2 & p_3(1-q_2) & (1-p_3)q_2 & (1-p_3)(1-q_2)\\
		p_4q_4 & p_4(1-q_4) & (1-p_4)q_4 & (1-p_4)(1-q_4)
	\end{bmatrix}	
\label{eq:MarkovMatrix}
\end{equation}
The transition matrix gives the probability of being in a specific state at any point of the Markov chain: let $\vec{\mu_1}$ be a column vector giving the probability distribution over the four states in the first iteration (which would be given by unconditional probabilities for $X$ and $Y$ in the first stage game), i.e\@. $\vec{\mu_1}^\T \vec{1}=1$ and e.g\@. ${\mu_1}_2=P(ud)$. Then the probabilities of being in each of the states $(\aClockWise)$ two iterations later, is 
$\vec{\mu_3}^\T=\vec{\mu_1}^\T \mat{M}^2$ 
with each ${\mu_3}_k$ giving the probability for the corresponding state $\text{xy} \in \mathbb{S}$
\cite[thm 2.3.2]{KemenyMarkov}.

It can be shown%
\footnote{\label{fn:Convergence}%
          For some values of $\vec{p}$ and $\vec{q}$, $\mat{M}$ will not satisfy the conditions for regular Markov chains \cite[pp. 36ff.]{KemenyMarkov}. However, by calculating $\mat{A}_4 := 1/4 \sum_{m=0}^3{\mat{M}^m}$ it can be shown that the chain converges nonetheless (cf. \cref{app:MatrixMult}). This suffices to prove that the average time spent in any of the four possible states will converge, and a unique stationary distribution will exist \cite[§2.2.2]{StroockMarkov}.
          }
that for general $\vec{p}$ and $\vec{q}$ and for any starting distribution $\vec{\mu_1}$, the Markov process given by $\mat{M}$ will converge to the same probability distribution (the stationary distribution) for later iterations. This stationary distribution is characterized by 
$\vec{\pi}^\T \mat{M}=\vec{\pi}^\T$ and $\vec{\pi}^\T \vec{1}=1$
\cite[thm 4.1.6]{KemenyMarkov}
($\Leftrightarrow$ $\vec{\pi}^\T$ is a left eigenvector of $\mat{M}$ with eigenvalue 1): this can be understood by recalling that the Markov chain is a process that repeatedly visits a finite set of states (in our case, $\mathbb{S}$). When such a process is infinitely long, it is intuitive that the average number of visits for each state will converge to a distribution over states that replicates itself. Mathematical analysis shows this distribution to be unique.

Since our game is infinitely iterated, $\vec{\pi}$ will give the expected distribution \cite[pp. 26f.]{StroockMarkov} over game outcomes: we know that in some future iteration, the distribution over $\mathbb{S}$ will reach $\vec{\pi}$, and that from then on it will stay there forever, making it the expected distribution for the total infinite chain. The law of large numbers for Markov chains proves that as the game proceeds over an infinite number of stages, the average of observed game outcomes will also converge to $\vec{\pi}$ \cite[§4.2]{KemenyMarkov}.

Since the average payoffs for $X$ and $Y$ depend only on this history of game outcomes, the expected average payoff for both players will also converge in an infinitely iterated game. Let $\bar{s}^{(t)}_i$ be the average score after $t$ iterations for player $i$. Let further $\pi^{(t)}_i$ be the expected value of $\bar{s}^{(t)}_i$. Now, as $t\rightarrow\infty$, the expected average payoffs converge to $\pi^{(\infty)}_i:=E\left[\bar{s}^{(\infty)}_i\right]=\vec{\pi}^\T \vec{s_i}$, which is the expected average payoff of the infinitely iterated game. Every stage $t>1$ of the infinitely iterated game is exactly the same (one historic move is common knowledge, the same stage game is played, and an infinite series of the same game follows), and every sequence of future games is probabilistic in nature (the strategies are probabilities): players who want to maximize their average stage game payoff over all games to come (as assumed here) will be interested in this expected average payoff $\pi^{(\infty)}_i$.%
\footnote{\label{fn:RiskNeutral}%
          It is noted that using expected payoffs generally implies that players are risk-neutral, an assumption made in order not to further complicate our analysis: a partial remedy may be to modify stage game payoffs to reflect risk aversion.
          }
It is therefore convenient to define $\pi_i := \pi^{(\infty)}_i$ from here on (which is directly comparable to stage game payoffs, as noted above).

\PrDy show that for any Markov matrix as given by $\mat{M}$, the expected average payoffs are given by%

\[
    \pi_i=\vec{\pi}^\T \vec{s_i}=\frac{D(\vec{p}, \vec{q}, \vec{s_i})}{D(\vec{p}, \vec{q}, \vec{1})}
    \text{, where}
\]
\vskip-3em
\begin{equation}
    \ \ \ \ 
    D(\vec{p}, \vec{q}, \vec{f}) := \det
    \begin{bmatrix}
        p_1 q_1 - 1 & p_1 - 1 & q_1 - 1 & f_1\\
        p_2 q_3     & p_2 - 1 & q_3     & f_2\\
        p_3 q_2     & p_3     & q_2 - 1 & f_3\\
        p_4 q_4     & p_4     & q_4     & f_4
    \end{bmatrix}
    \equiv \lambda * \vec{\pi}^\T \vec{f}
\label{eq:ExpectedDeterminant}
\end{equation}
(with $\lambda \in \mathbb{R}$, and $\vec{f} \in \mathbb{R}^4$. The denominator cancels out the scale factor $\lambda$). Then, because $D(\vec{p}, \vec{q}, \vec{f})$ is a linear function in $\vec{f}$, for any $\alpha, \beta, \gamma \in \mathbb{R}$%

\begin{equation}
      \alpha\pi_X + \beta\pi_Y + \gamma = 
      \frac{D(\vec{p}, \vec{q}, \alpha\vec{s_X} + \beta\vec{s_Y} + \gamma\vec{1})}
      {D(\vec{p}, \vec{q}, \vec{1})}
\label{eq:LinearCombinationScores}
\end{equation}
This is the core of ZD strategies: since by choice of their strategy $X$ and $Y$ both independently control one column of the determinant in \cref{eq:ExpectedDeterminant}, they can unilaterally choose a strategy that makes their respective column linearly dependent on $\vec{f}$. In this case the determinant in the numerator of \cref{eq:LinearCombinationScores} is 0 (hence \emph{``zero determinant''} strategies):%
\footnote{%
          Of course, this leaves $\pi_i$ undefined when $\vec{p}$ or $\vec{q}$ are $(1, 1, 0, 0)^\T$ (when the denominator is 0). In such cases, the Markov chain may be analyzed on a reduced subset of $\mathbb{S}$ (see \cref{app:MatrixMult}).
          }
$X$ can enforce the linear relationship $\alpha\pi_X + \beta\pi_Y + \gamma = 0$ to hold by choosing her strategy such that it satisfies the following constraint where $\alpha, \beta, \gamma$ are chosen by her freely such that $\vec{p}$ remains in the realm of probability vectors (i.e\@. for every element $k$ of the strategy $0 \leq p_k \leq 1$):%

\begin{equation}
	\vec{p} = 
	  \alpha\vec{s_X} + \beta\vec{s_Y} + \gamma\vec{1} + (1, 1, 0, 0)^\T
\label{eq:GeneralZDstrat}
\end{equation}
All following analyses will take the illustrative viewpoint of $X$: $Y$ has the exact same options to play zero determinant strategies with $(q_1, q_3, q_2, q_4)^\T = \alpha\vec{s_X} + \beta\vec{s_Y} + \gamma\vec{1} + (1, 0, 1, 0)^\T$ (where structural differences in the equation are due to the ordering of payoff vectors).

	\section{Mischief and extortion}%
  \label{sec:MischiefZD}
%
\Cref{eq:LinearCombinationScores} ``allows much mischief'' \cite[p. 2]{PrDy}: by choosing values $\alpha, \beta, \gamma$ that keep her strategy $\vec{p}$ as defined in \cref{eq:GeneralZDstrat} in the realm of possibility vectors, $X$ can unilaterally impose certain constraints on the iterated game's expected average scores:%
\footnote{%
          Note that \cref{eq:GeneralZDstrat} generally leaves one degree of freedom for all strategies discussed here: for all parameters discussed, a multitude of ZD strategies exists.
          }

Firstly, $X$ may choose to set $\alpha=0$, yielding $\pi_Y = -\gamma / \beta$. By doing so, she can unilaterally determine $Y$'s expected average payoff: $Y$'s strategy will have no impact whatsoever on his average payoffs in the long run, and he will thus be indifferent between all his strategy options. However, $X$'s own payoff will then fully depend on $Y$, and will often not be very high. Due to the (potentially) malicious character of such strategies, this class of strategies will be called \textit{ZDmischief} strategies in this paper. They have been studied before (with a different proof and nomenclature) by \cite{SigmundNowakEqual}.

The feasible payoffs to be set unilaterally depend on the payoff matrix; in the next section I present feasible values for the games analyzed here. It can already be noted, however, that it will be impossible to set scores which are less than the opponent's maximin payoff in any game: $Y$ can always revert to his maximin strategy to escape the mischief. Indeed, this lower bound is confirmed in my analytical derivation.

Secondly and most interestingly, $X$ may enforce a linear relation between her and $Y$'s scores: $X$ may ensure herself a multiple of every surplus $Y$ earns over a certain offset. By setting $\gamma = -(\alpha \Delta + \beta \Delta)$, for any offset $\Delta$, $X$ enforces $(\pi_X - \Delta) = \chi(\pi_Y - \Delta)$, where $\chi:=-\sfrac{\beta}{\alpha}$ is the ``extortion factor.'' For values $\chi > 1$ such strategies could be described as enforcing an ``unfair,''%
\footnote{%
          The normative implications of equality are not further discussed in this paper. It seems obvious that $\chi>1$ would be described as ``unfair'' by most observers in almost every possible application. When used, normative language will put in double inverted commas to indicate metaphorical use of the terms while applied to the theoretical phenomena discussed here.
          }
extortionate share of payoffs for $X$; this class of strategies with $\alpha \neq 0 \neq \beta$ will thus be called \textit{ZDextortion} strategies in this paper.

Generally the range of feasible extortion factors will depend on the choice of $\Delta$. In the games analyzed here, there are ranges for $\Delta$ in which there are upper bounds on the extortion factor, and ranges for $\Delta$ which do not limit extortion.

Natural values for $\Delta$ are maximin payoffs $r$: by setting $\Delta=r$, $X$ ensures that she will get $\chi$-times any surplus $Y$ earns on his maximin payoff. Since $Y$ cannot by himself earn more than $r$, it should be possible to extort any unequal share of payoffs above maximin (the extortion argument going somewhat like this: ``if you want to earn more than you could by yourself, you have to pay me a larger share for my cooperation''). Indeed, in all games analyzed here, maximin payoffs are a lower bound for $\Delta$ if there is to be no upper bound on $\chi$ (see the next section).

In some games it might further be tempting to set a smaller offset, e.g\@. $\Delta = 0$ to achieve $\pi_X = \chi\pi_Y$. Often, this does not yield feasible strategies. When it does, however, there will typically be an upper bound on the extortion factor. The next section discusses cases in which such strategies with $\Delta=0$ appear to be rather trivial.

Of course, both $X$ and $Y$ are free to play ZD strategies; if both players play \textit{ZDextortion} strategies with $\chi>1$, their scores will be the solution to both extortion constraints: in most games $\pi_X=\pi_Y=\Delta$.%
\footnote{%
          Let $\Delta_i$ denote the offset chosen by $i\in{X, Y}$, and assume $\chi_i>1$. Then a simple calculation to solve the two resulting ZD constraints shows that the player choosing the smaller offset will outperform the other. Thus, $\Delta_X = \Delta_Y$ will be a rational assumption when $\Delta$ has a lower bound. Then both constraints are satisfied when $\pi_X=\pi_Y=\Delta$, regardless of extortion factors.
          }

Lastly, $X$ might try to secure her own score independently of $Y$'s actions by setting $\beta=0$. Not surprisingly, this is not possible in most interesting games (such as the common ones covered here).%
\footnote{%
          A straightforward calculation following constraint \cref{eq:GeneralZDstrat} shows that in this case $\vec{p}$ is only in the realm of possibility vectors if one action is ``wantlessly dominant,'' i.e\@. all possible outcomes when choosing one action are better than all other possible outcomes when playing the other ($s_{X_{1,2}} \ll s_{X_{3,4}}$, or vice versa). In that case, $X$ could choose a fixed $\pi_X$ to achieve in the long run by mixing between the dominant and the dominated strategy---neither such a strategy nor games with such strongly dominant strategies seem particularly interesting, since playing anything but the dominant strategy would seem irrational.         
          }

	\section{Exemplary ZD strategies in 2x2 games}
  \label{sec:ZDparameters}

Possible parameters for ZD strategies in the four games exemplified here are summarized in \cref{tab:ZDStrats}, where $\tilde{\pi}_Y$ denotes $Y$'s expected average payoff to be enforced by $X$'s \textit{ZDmischief} strategy, and $\Delta$ and $\chi$ are used as parameters of \textit{ZDextortion} strategies as previously introduced. The table also gives one numeric example for each strategy in each game, with payoffs as introduced in \cref{IntroStageGames}. For the example of an infinitely iterated Game of Chicken, exemplary and average gameplay of the derived strategies is also presented. The derivation of feasible parameters can be found in \cref{app:ZDstratDerivation}, where ranges are provided for a general symmetric game and a general iBS.

  \subsection{Feasible parameters}

\begin{figure}[hbt]
  \centering
  \def\svgwidth{10cm}
  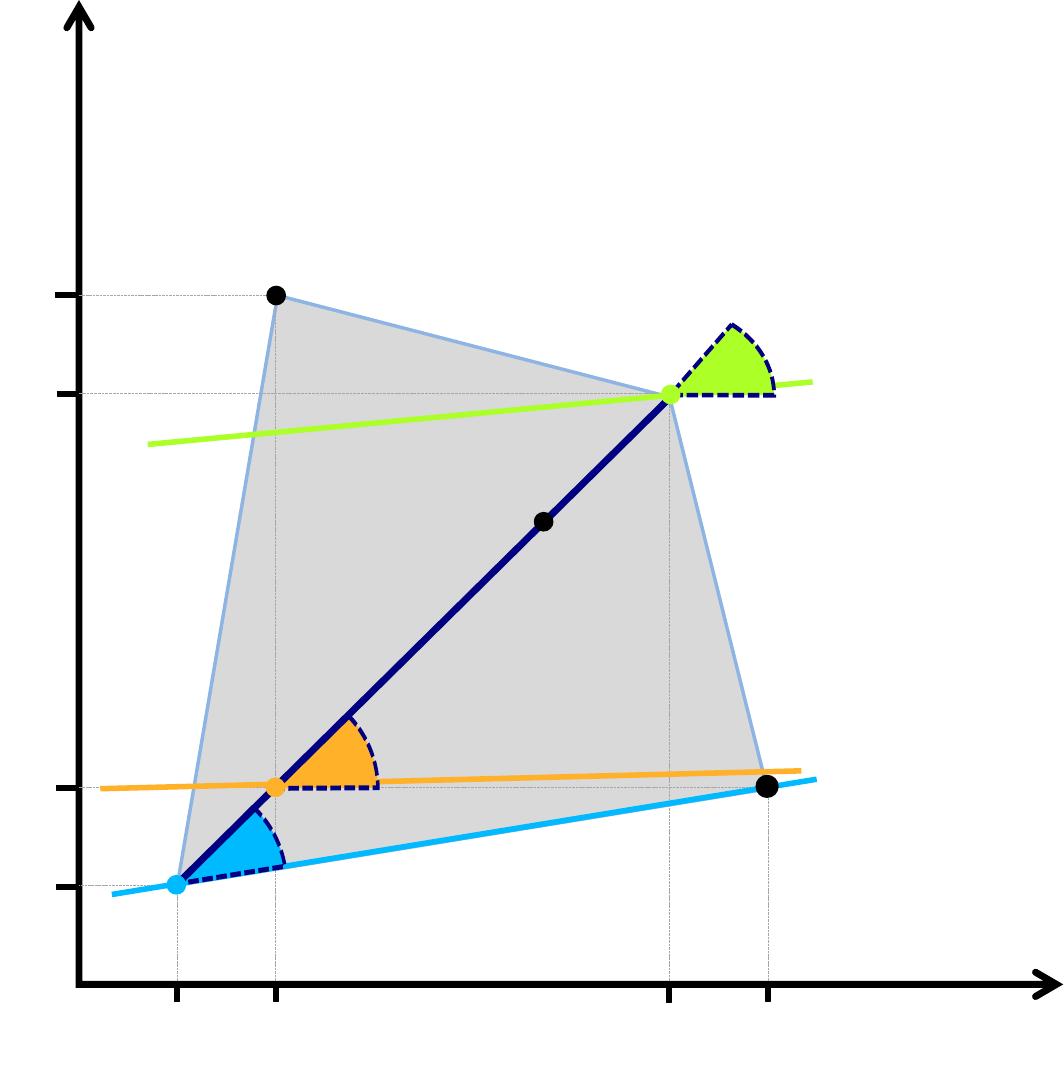
  \caption{Graphical representation of feasible \textit{ZDextortion} strategies for $X$}
	\label{fig:ZDConstraints}
  \floatfoot{Note: Three kinds of \textit{ZDextortion} strategies are illustrated: the dark blue line represents the feasible range for offsets $\Delta$. The colored dots give three values for $\Delta$ along with feasible ranges for the extortion factor $\chi$ (represented by the colored arcs) and the resulting constraints for one particular value of $\chi$ (the lines). When $\Delta=r$ (orange), $\chi$ may take any value from $[1, \infty[$, enabling the ZD constraint to be almost flat. When a smaller $\Delta$ is chosen, the set of feasible payoffs imposes an upper bound on $\chi$, and in the extreme case of $\Delta=P$ (light blue), $X$'s strategy is equivalent to always playing $\aDOWN$. Finally, $\Delta$ is shown at its upper limit value $\Delta=R$ (green). Asterisks mark stage game Nash equilibria.}
\end{figure}

``Fair'' extortion strategies with $\chi=1$ exist for all games analyzed here (and $\chi=1$ implies that $\Delta$ is canceled out). In \cref{app:ZDstratDerivation} it is shown that such strategies exist in all quantitatively symmetric and in general Battle of the Sexes games, and it is clear that they can be construed by adapting parameters in ordinally symmetric games (where the ZD constraint would then intersect the payoffs from $(\aUP, \aUP)$ and $(\aDOWN, \aDOWN)$, cf. \cref{fig:ZDConstraints}). This result is in line with more general theorems proved for a \mem{1} imitate-the-best strategy in \cite{Imitation}. \PrDy note that tit-for-tat is a special case of \textit{ZDextortion} strategies in symmetric games with $\chi=1$: $\aTFT$ is $\vec{p} = (1, 0, 1, 0)^\T$, one of the limit values of \cref{eq:GeneralZDstrat} \cite[p. 3]{PrDy}.

In symmetric games, and particularly in those three analyzed here, a mischievous $X$ may unilaterally force $\pi_Y$ to any value between $Y$'s pure strategy maximin payoff ($r=\max \{S, P\}$) and the lower of his two highest payoffs ($s:=\min \{R, T\}$).

For \textit{ZDextortion} strategies in symmetric games, there is no upper bound on the extortion factor $\chi$ when adequate $s \geq \Delta \geq r$ is chosen. Therefore $X$ may decide to skew virtually all surplus payoff on the maximin payoff from $Y$. Choosing an offset $\Delta$ smaller than the maximin payoff is possible if and only if $P \leq \Delta < r$: then $\chi$ has an upper bound. This is the case in iGC as defined here (where $r = S > P$). Thus, for $\Delta=0$ to be possible, $P \leq 0$ must hold, and if $r>0$, $\chi$ will have upper bounds which will then give the edges of the set of feasible payoffs.

For illustration, assume the case of limit values, i.e\@. $\Delta=P<S$ and $\chi=(T-\Delta)/(S-\Delta)$. Then the resulting ZD constraint is rather trivial: it is the lower edge of the set of feasible payoffs (the line $(\aDOWN, \aDOWN) \longrightarrow (\aDOWN, \aUP)$ in \cref{fig:ZDConstraints}), the same one which could be obtained by simply always playing $\aDOWN$. Indeed, the resulting \textit{ZDextortion} strategy with these limit parameters will play $\aUP$ for a finite number of iterations and will then continue to play $\aDOWN$ forever ($\vec{p} = (p_1, p_2, 0, 0)^\T$, with $p_2 < p_1 < 1$). In \cref{fig:ZDConstraints}, I give a graphical interpretation of feasible parameters for extortion strategies.

In Battle of the Sexes games, no \textit{ZDmischief} strategies exist, and extortion is generally limited by upper and lower bounds.%
\footnote{%
          This may provide some insights into why $X$ and $Y$ continue to be fascinated by them. Romantically, $\chi<1$ is possible in iBS games, unlike in symmetric games.
          }
\textit{ZDextortion} strategies with $\chi>1$ only exist in iBS games if $C=L$; they have the same characteristic highlighted above for iGC: if they exist at all, they will give the edges of the set of feasible payoffs in their limit values.

\begin{sidewaystable}[phbt]%
\centering
\begin{tabular}[t]{lcccc}
      \toprule
          & iPD & iSH & iGC & iBS \\
      \midrule
      \textit{ZDmischief} & 
          $P \leq \tilde{\pi}_Y \leq R$ &
          $P \leq \tilde{\pi}_Y \leq T$ &
          $S \leq \tilde{\pi}_Y \leq R$ &
          no feasible values     
      \vspace{5pt} \\      
      \cmidrule{2-5}
      \ \ \ ex. & 
           \pbox{20cm}{\small $\vec{p} = (0.8, 0.6, 0.1, 0)^\T$ \\ $\Rightarrow \pi_Y = 1$}& 
           \pbox{20cm}{\small $\vec{p} = (0.8, 1, 0.8, 0)^\T$ \\ $\Rightarrow \pi_Y = 8$}& 
           \pbox{20cm}{\small $\vec{p} = (0.65, 0.55, 0.05, 0.25)^\T$ \\ $\Rightarrow \pi_Y = 2.5$}& 
           n/a  
      \\ 

      \midrule
      
      \textit{ZDextortion:} \vspace{5pt}  \\
      $\chi<1$   \vspace{5pt} & n/a & n/a & n/a & 
        \parbox{2.5cm}{\centering $L = \Delta = C$ \\ $\chi \geq \frac{\Delta-D}{\Delta-F}$}  \\
      \cmidrule{2-5}
      $\chi=1$   \vspace{5pt} & always possible & always possible & always possible & always possible \\
      \cmidrule{2-5}
      $1<\chi$ \vspace{5pt} & 
        $\Delta \in [P, R]$ & 
        $\Delta \in [P, T]$ & 
        $\Delta \in [S, R]$ &
        \parbox{2.5cm}{\centering $L = \Delta = C$ \\ $\chi \leq \frac{F-\Delta}{D-\Delta}$} \\

      \cmidrule{2-5}
      $\Delta < r$ \vspace{5pt} & n/a & n/a &
                  $1 \leq \chi \leq 
                  \frac{T - \Delta}{S - \Delta}$
                  & n/a 
      \\
      
      \cmidrule{2-5}
      
      \ \ \ ex. & 
           \pbox{20cm}{\small $\vec{p} = (0.64, 0.18, 0.28, 0)^\T$ \\ $\Rightarrow (\pi_X - 1) = 10(\pi_Y-1)$}& 
           \pbox{20cm}{\small $\vec{p} = (0.82, 0.92, 0.8, 0)^\T$ \\ $\Rightarrow (\pi_X - 8) = 10(\pi_Y-8)$}& 
           \pbox{20cm}{\small $\vec{p} = (0.28, 0, 0.1, 0.36)^\T$ \\ $\Rightarrow (\pi_X - 2) = 10(\pi_Y-2)$}& 
           \pbox{20cm}{\small $\vec{p} = (1, 1, 0, 0.6)^\T$ \\ $\Rightarrow (\pi_X - 1) = 2(\pi_Y-1)$} 
      \\ 
      
      \bottomrule
\end{tabular}
\caption{Possible ZD strategies in common 2x2 games}
\label{tab:ZDStrats}
\end{sidewaystable}

  \FloatBarrier
  \subsection{Typical gameplay}

\begin{figure}[hp]
  \centering
  \def\svgwidth{15cm}
  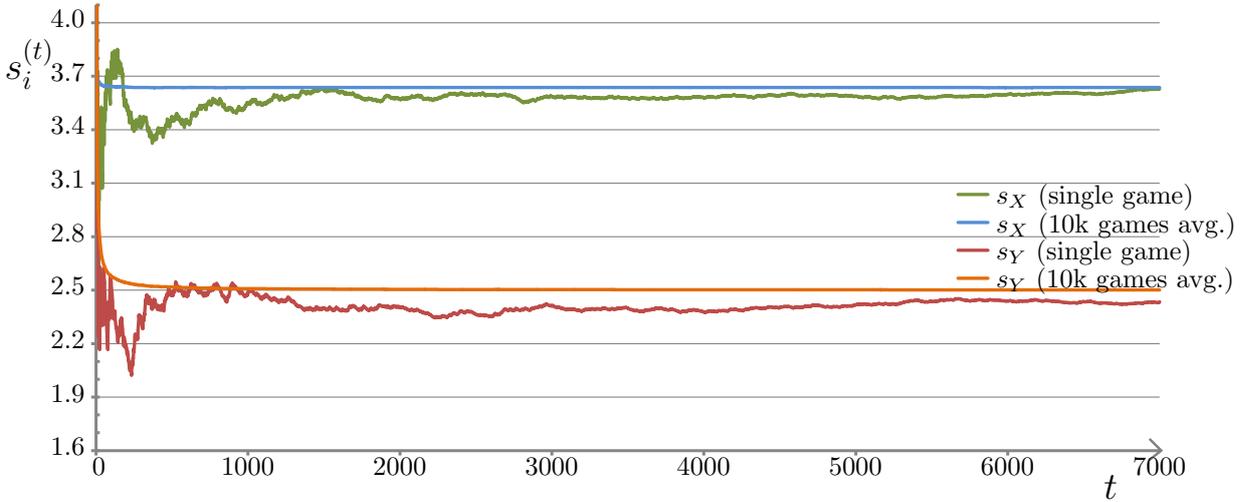
  \caption{iGC games: \textit{ZDmischief} with $\tilde{\pi}_Y = 2.5$ vs. randomizer}
	\label{fig:iGC_MisVsRandom}
  \floatfoot{Note: single game: $(s^{(7000)}_X, s^{(7000)}_Y) \approx (3.63,	2.43)$. On average over 10,000 games, they are $(3.64, 2.50)$.}
\end{figure}

\begin{figure}[hp]
  \centering
  \def\svgwidth{15cm}
  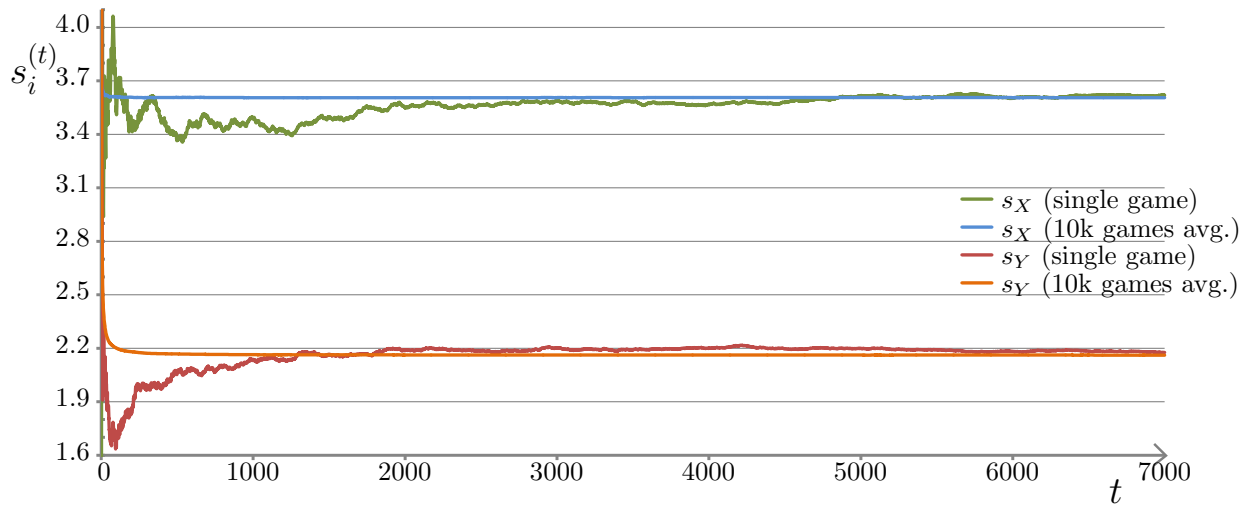
  \caption{iGC games: \textit{ZDextortion} with $\chi=10, \Delta=2$ vs. randomizer}
	\label{fig:iGC_ExtVsRandom}
  \floatfoot{Note: single game: $(s^{(7000)}_X, s^{(7000)}_Y) \approx (3.62,	2.18)$. On average over 10,000 games, they are $(3.61,	2.16)$.}
\end{figure}

\begin{figure}[hp]
  \centering
  \def\svgwidth{15cm}
  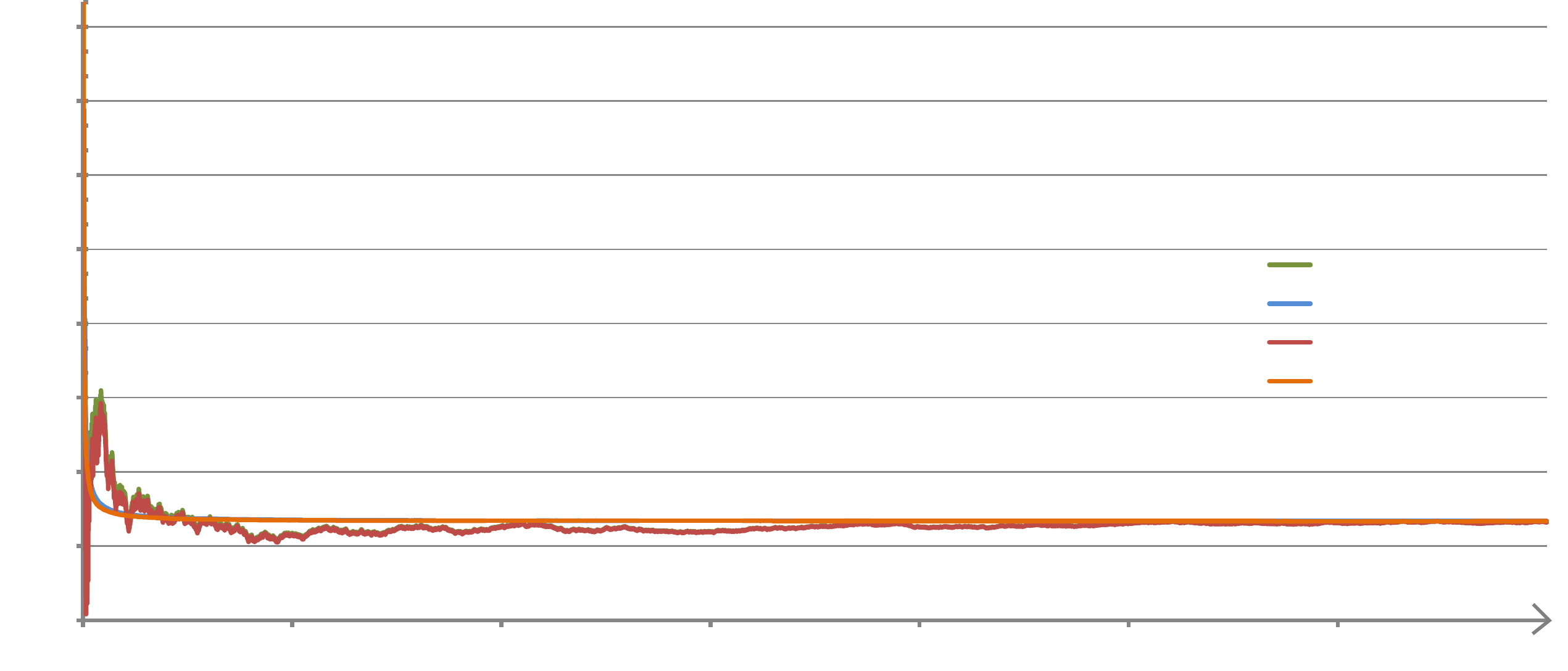
  \caption{iGC games: \textit{ZDextortion} with $\chi=10, \Delta=2$ vs. $\aTFT$}
	\label{fig:iGC_ExtVsTFT}
  \floatfoot{Note: single game: $(s^{(7000)}_X, s^{(7000)}_Y) \approx (2.00,	2.00)$. On average over 10,000 games, they are $(2.00,	2.00)$.}
\end{figure}

ZD strategies typically ``work'' very fast, yielding average scores that are psychologically ``close enough'' to the expected values within a few hundred iterations of the game.%
\footnote{%
          An interactive javascript-website impressively demonstrates this by offering a simple iPD game to play \cite{ZDwebgame}.
          }
\Cref{fig:iGC_MisVsRandom,fig:iGC_ExtVsRandom,fig:iGC_ExtVsTFT} present exemplary gameplay in iGC:%
\footnote{%
          The simulation code used to obtain this data can be found in \cref{app:ZDsimulation}.
          }
one single game is compared to an average over 10,000 games to show typical and average convergence of payoffs. All panels show games in which $X$ plays a numerical ZD strategy provided in \cref{tab:ZDStrats}. In the first two panels, they play a randomizer strategy $Y$ which plays each move with equal probability, i.e\@. $\vec{q}=(0.5,0.5,0.5,0.5)^\T$. It should be noted that convergence against the randomizer strategy is relatively slow, compared to other strategies (see \cref{sec:ZDConvergence} below).

The third panel shows the extortionate strategy $X$ against tit-for-tat $Y$, which in effect is a game between two \textit{ZDextortion} strategies, one with $\Delta=2, \chi=10$ and one with $\chi=1$. Both resulting constraints are satisfied for the expected average payoffs of $\pi_X=\pi_Y=2$. It is interesting to note that the extortioner $X$ has a slightly higher average score in the first few hundred iterations, which stems from her initiation of defection against the cooperative $\aTFT$. However, this small advantage is quickly averaged over as more iterations are played. This example demonstrates a more general truth: two extortioners with $\chi>1$ competing against each other will generally earn very mediocre payoffs, as a simple calculation following the two ZD constraints shows.

All panels demonstrate that iterated games between probabilistic strategies are sto\-chastic processes: every single iterated game has entirely different gameplay, and in (very unlikely) singular cases gameplay may deviate completely from the examples shown here: it is possible, if not likely, that $(\aUP, \aUP)$ is played for any finite duration of gameplay. The average over many games shows a very strict convergence, however. This is further discussed in the following section.

	\section{Convergence speed of ZD strategies}
  \label{sec:ZDConvergence}

The existence and behavior of ZD strategies has been proven and analyzed for \emph{infinitely} iterated 2x2 games; regardless of whether \textit{ZDmischief} or \textit{ZDextortion} strategies are played, the expected average payoffs of the infinitely iterated game, $\pi_X$ and $\pi_Y$, are given by \cref{eq:ExpectedDeterminant} as a function of $\vec{p}$ and $\vec{q}$, both players' \mem{1} strategies (or $Y$'s equivalent \mem{1} strategy, if he plays a longer-memory strategy).

However, even with the rationalization discussed earlier (i.e\@. no player ever certainly knows a single iteration to be the last one), human actors in economic applications of the model will be interested in how quickly they may expect to see the average payoffs converge: in what time frame may one expect the ZD constraints to be ``almost satisfied''? While it has been shown analytically that playing a ZD strategy will guarantee the intended linear constraint in the (very) long run, it is interesting to ask how close to the limit value of the payoff convergence a player can get within finite time spans.%
\footnote{%
          It is noted that in terms of classical game theory the game must be ``infinitely iterated'' at least in so far that the players do not know when it will end. If one iteration can be made out to be the last, always playing $\aDOWN$ follows as only rational strategy from backwards induction. This phenomenon is known as the Chainstore paradox \cite[pp. 128ff.]{RasmusenGT}.
          }

For this reason, the convergence speed of ZD strategies is of interest: how quickly can one assume the average payoffs to be within a reasonable neighborhood of the limit in the infinitely iterated game? Since the convergence of ZD strategies is due to the convergence of Markov chains, this question can be answered by Markov chain theory.

The mean ergodic theorem \cite[§2.3.1]{StroockMarkov} can be applied to show that (with notation as introduced above)

\begin{equation}
      E\left[ (\bar{s}^{(t)}_i - \pi^{(\infty)}_i)^2 \right] \leq \frac{C}{t} 
\label{eq:ConvergenceBoundary}
\end{equation}
for $i \in {X, Y}$, where $C$ is a constant determined by the Markov transition matrix (and thus by $\vec{p}$ and $\vec{q}$) and the payoff vectors. To be precise, knowing that $\mat{A}_4$ has at least one column of non-zero elements (see \cref{app:MatrixMult})

\begin{equation}
C = 6 * \hat{s}_i^2 * \epsilon^{-1}
\end{equation}
Here, $\hat{s}_i$ denotes the largest possible stage game payoff for player $i$, and $\epsilon > 0$ is the minimum element of any column in $\mat{A}_4$. For example, in iGC between the \textit{ZDextortion} strategy provided in \cref{tab:ZDStrats} and $\aAllU$ (the strategy always playing $\aUP$), $C \approx 645$; the same strategy against randomizer yields $C \approx 1333$: it is clear that this theoretic and general result is too broad an approximation to impress ``human actors'' (very roughly speaking, these values imply that the average scores after 1000 iterations can be expected to be within $\pm 1$ of the expected average scores of the infinitely repeated game).

Additionally, no statement can be made stronger than the expected value of the deviation: when $\vec{p}$ and $\vec{q}$ are kept general and $t<\infty$, $\bar{s}^{(t)}_i$ can theoretically take any value from the convex set of feasible payoffs; deviation from the expected value $\pi_i$ (the convergence limit) is just increasingly unlikely for large $t$. Nonetheless, the strong law of large numbers guarantees that in an average over a large number of independent iterated games, $(\bar{s}^{(t)}_i - \pi^{(\infty)}_i)^2 \leq \frac{C}{t}$ will hold. This result over a large number of games can be observed in \cref{fig:iGC_ExtVsAllD,fig:iGC_ExtVsRandom,fig:iGC_ExtVsTFT,fig:iGC_MisVsRandom}, where the average payoffs over 10,000 games show a strong continuous convergence---the simulations also suggest that convergence is \emph{much} faster than implied by the general results: maybe this data will bear more attractiveness for real-world players.

Lastly, since the stationary distribution $\vec{\pi}$ of the Markov chain depends on the strategies $\vec{p}$ and $\vec{q}$ being fixed, it could be assumed that by continuously changing his strategy $\vec{q}$, $Y$ may actually evade the convergence of payoffs: by ``playing inside the equilibration timescale,'' $Y$ might try to ``keep the game out of Markov equilibrium'' \cite[p. 4]{PrDy}. An argument similar to that in \cref{sec:Generality} shows that this is impossible: any variation of strategies, rapid as it may be, can be averaged over to obtain an equivalent, fixed \mem{1} strategy, which suffices for the Markov chain to converge. \PrDy present a formal proof of this idea \cite[appx B]{PrDy}. 

However, the results derived in this section highlight another aspect: the convergence speed of the Markov chain depends on both $X$'s and $Y$'s strategy choice. If the game continues for infinitely long, this will be irrelevant, since $\pi_i$ does not depend on convergence speed; if the players have time preferences, or if the infinite game has a positive probability of ending in every iteration (signified by $\delta<1$), however, they may find retardation of convergence attractive in any iteration $t<\infty$. This result is further discussed in the following section.

	\section{Best responses to ZD strategies}
  \label{sec:BestResponse}

If $X$ is to play a ZD strategy, what is the best reaction for $Y$? Under the assumptions made, $Y$ would be interested in $\vec{\bar{q}}:=\arg\max_{\vec{q}} \pi_Y(\vec{p}, \vec{q})$. Recall that this implies that $\delta=1$, i.e\@. $Y$ has no time preferences: he values future payoffs exactly as much as present ones (and therefore maximizes the simple average over all iterations). Of course, this is a strong assumption; for this section to be formally correct, we will now also consider time preferences, i.e\@. $\delta<1$.%
\footnote{%
          We will not relax the assumption of risk neutrality (see \cref{fn:RiskNeutral}).
          }

A comparable measure to $\pi_i=\pi^{(\infty)}_i$ as previously used is given by the expected average payoff over all stage games to come, weighted by time preference \cite[pp. 210f.]{EichbergerGT}, defined as (cf. \cref{eq:DiscountedPayoff})

\begin{equation}
\hat{\mathscr{P}}_i (\vec{p}, \vec{q}) := (1-\delta) * \sum^{\infty}_{t=1}{ \delta^{t-1} E\left[ \widehat{\pi_{i, t}}(\vec{p}, \vec{q}) \right] }
                    \widehat{=} \left( \sum_{t=1}^{\infty}{\delta^{t-1}} \right)^{-1} * E \left[\mathscr{P}_i(\sigma_X, \sigma_{Y}) \right]
\end{equation}
where notation is simplified by defining a random variable $\widehat{\pi_{i, t}}(\vec{p}, \vec{q})$ as ``player $i$'s stage game payoff in iteration $t$, given both players' probabilistic strategies.''

Thus, if time-indifferent, $Y$ will try to choose $\vec{\bar{q}}$. If he has time preferences, $Y$ will try to choose $\vec{\hat{q}}:=\arg\max_{\vec{q}} \hat{\mathscr{P}}_Y(\vec{p}, \vec{q})$. It should be noted that for $\delta$ small enough, our analysis becomes irrelevant: when $Y$ does not care much about the implications his action in iteration $t$ has in iteration $t+1$ (viz\@. impacting $X$'s future choice of action through her \mem{1} strategy $\vec{p}$), the infinitely iterated game effectively becomes an infinite series of unconnected stage games. I thus assume $\delta$ to be in a range where ZD analysis is still relevant, and focus on the implications of $\delta<1$. With these assumptions, best response strategies $\vec{\bar{q}}$ and $\vec{\hat{q}}$, and other responses that might be deemed ``best'' are now discussed for different ZD strategies.

In the case of \textit{ZDmischief} strategies, $Y$'s expected average score $\pi_Y$ will be fixed to $\tilde{\pi}_Y$ regardless of his strategy choice (including, as shown in \cref{sec:Generality}, more elaborate strategies): there simply is \emph{no} best response \textit{per se}. $Y$ might choose to deliberately keep $\pi_X$ low, hoping to make $X$ stop her mischief---but by itself this will have no impact on $Y$'s score whatsoever. If $Y$ has time preferences, however, the results from \cref{sec:ZDConvergence} imply that by choosing an appropriate strategy $\vec{\hat{q}}$, $Y$ is able to slow convergence down. Since he values present payoffs higher than distant ones (and since the value of very distant payoffs is $\lim_{t \rightarrow \infty} \delta^{t-1} \widehat{\pi_{i, t}} = 0$), such strategies may be best responses for him---the concrete answer depending on $\delta$, $\vec{p}$, and $\vec{s_Y}$. Thus, for \textit{ZDmischief} strategies, best responses will exist for $\delta<1$. Otherwise $Y$ will be indifferent between all strategies.

Now assume that $X$ plays a \textit{ZDextortion} strategy with small $\Delta$ and $\chi \geq 1$. Then $Y$'s best response $\vec{\bar{q}}$ ensures that $X$ will get a multiple: since $X$ imposes the constraint $(\pi_X - \Delta) = \chi(\pi_Y - \Delta)$, $Y$'s maximization of his expected average score makes sure that $X$ will be better off. In all games analyzed here, this leads $Y$ to always play $\aUP$ ($\aAllU$), yielding a payoff distribution on the right-hand edge of the set of feasible payoffs in \cref{fig:ZDConstraints}: by following nothing but his private interests, $Y$ can only play in the hands of the extortioner. Regardless of how large $\chi<\infty$ is, $Y$ will be better off making his extortioner rich than keeping both players poor. 

If $Y$ has time preferences, two cases are possible: if $\delta$ is large enough, the dynamic leading $Y$ to fully committing to his extortioner (e.g\@. by playing $\aAllU$) will work when $Y$ maximizes $\hat{\mathscr{P}}_Y$ as well. When $\delta$ is too small, the promise of gaining $\pi_Y>r$ over many future iterations will not suffice, and $Y$ should revert to more short-term strategies. That is, $\delta<1$ puts another constraint on the maximum extortion factor, if $X$ intends to successfully extort $Y$. Of course, the above remark on the possibility (and desirability) of convergence retarding strategies applies to \textit{ZDextortion} strategies as well.

Naturally, giving in to $X$'s mischief or extortion will disappoint some players, even if it presents the best response by private interest. Suppose, for example, $Y$ exhibits a sense of justice (or social preferences) and refuses to contribute to $X$'s ``unfair'' extortion: by defecting (or playing a ZD strategy himself), he can enforce the ``fair'' but mediocre payoffs $\pi_X=\pi_Y=\Delta$. Thus, ZD strategies in 2x2 games resemble an iterated ultimatum game, where (any) one player might be tempted to enforce an unequal distribution of payoffs, and the other is left with the decision to either comply or sacrifice own payoff. Similarly to the ultimatum game, the standard best response is to accept any ``immoral'' offer that is at least slightly better than the maximin payoff, despite a multitude of empirical observations confirming experiment subjects to tend to more equal splits (cf. \cite[pp. 358ff.]{RasmusenGT}. See also \cite{OechsslerUltimatum} for a description of circumstances that may drive $X$ towards a more equal split by herself).

An important difference to the original ultimatum game is that in playing against a \textit{ZDextortion} strategy with $\Delta=r, \chi>1$ there exist so far unstudied responses which present a stronger retaliation than merely falling back to maximin payoffs: if any stage game strategy combination exists yielding lower payoffs than maximin for both players, $Y$ can choose a strategy such that $\pi_Y<\Delta$. In this case, the extortionist's constraint on expected scores, $(\pi_X - \Delta) = \chi(\pi_Y - \Delta)$, backfires, as he will then receive \emph{less} than $Y$. Thus, the extortionist is herself liable to extortion.

\begin{figure}[hbtp]
  \centering
  \def\svgwidth{15cm}
  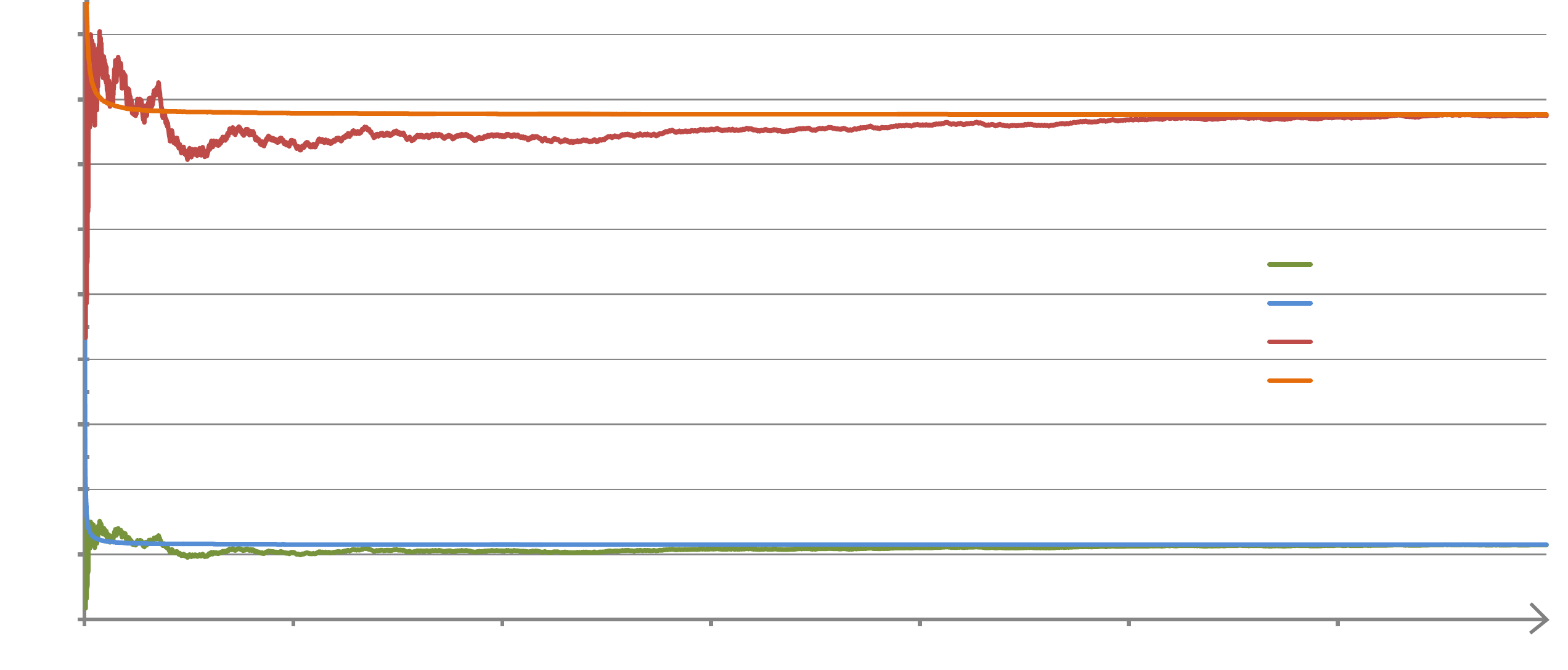
  \caption{iGC games: \textit{ZDextortion} with $\chi=10, \Delta=2$ vs. $\aAllD$}
	\label{fig:iGC_ExtVsAllD}
  \floatfoot{Note: single game: $(s^{(7000)}_X, s^{(7000)}_Y) \approx (0.53,	1.85)$. On average over 10,000 games, they are $(0.53,	1.85)$.}
\end{figure}

Such retaliation depends on the possibility of achieving $\pi_Y<\Delta$. For iGC as defined here, an example of retaliation is shown in \cref{fig:iGC_ExtVsAllD}, where the \textit{ZDextortion} strategy provided above plays against $\aAllD$ (the strategy always playing $\aDOWN$), and the ZD constraint is satisfied by $\pi_X<\pi_Y<\Delta=r$. However, a savvy extortionist can anticipate this reaction and avoid choosing $\Delta=r$, settling for $\Delta=P$ and an upper constraint on his extortion.

Retaliation is feasible whenever $\frac{D(\vec{p}_{ZD}, \vec{q}, \vec{s_Y})}{D(\vec{p}_{ZD}, \vec{q}, \vec{1})} = \pi_Y < \Delta$ yields a valid probability vector $\vec{q}$. Knowing that $\Delta$ is bounded below in $P$ for symmetric games, the feasibility of retaliation then depends on $S$ being relatively small: if and only if $(T+S) < 2P$, there exist payoff combinations lower than $P$, viz\@. combinations between $(\aDOWN, \aUP)$ and $(\aUP, \aDOWN)$. In such cases, even the extortion strategy with the lowest possible offset $\Delta=P$ is itself subject to extortion.

While the con\-ven\-tion\-ally-valued iPD game analyzed by \PrDy does not allow this, differently valued iPD games do: if $S$ is relatively small, $(T+S) < 2P$ holds.%
\footnote{%
          Take, for example, the modified iPD game with payoffs $(R,S,T,P)_{mPD}=(10,0,11,9)$. In this game, any \textit{ZDextortion} strategy will be outperformed by the randomizer strategy. E.g., for $\vec{p}=(0.91, 0.71, 0.92, 0)^\T$, an extortioner with $\Delta=P=9, \chi=10$, the expected average scores are $\pi_X \approx 6.46$, and $\pi_Y \approx 8.75$.
          }
Of course, this can also hold in iSH games. In iBS games, such retaliation is not possible, since $(F+D) > 2C$ is implied by definition.

Interestingly, in a simulated tournament of several well-known strategies in the con\-ven\-tion\-ally-valued iPD, Stewart and Plotkin report a strategy obtained by setting $\Delta=R$ to be particularly successful in terms of average score \cite{StewartPlotkin}. This strategy effectively vows to suffer more from every deviation from mutual cooperation than its opponent: it can easily be exploited (by choosing a strategy to earn any average payoff $\pi_Y < R$). It can only be assumed that the success of Stewart and Plotkin's strategy depends heavily on the zoo of strategies playing.

	\section{ZD strategies and the Folk theorem}
  \label{sec:ZDFolk}

\begin{figure}[hbt]
  \centering
  \def\svgwidth{10cm}
  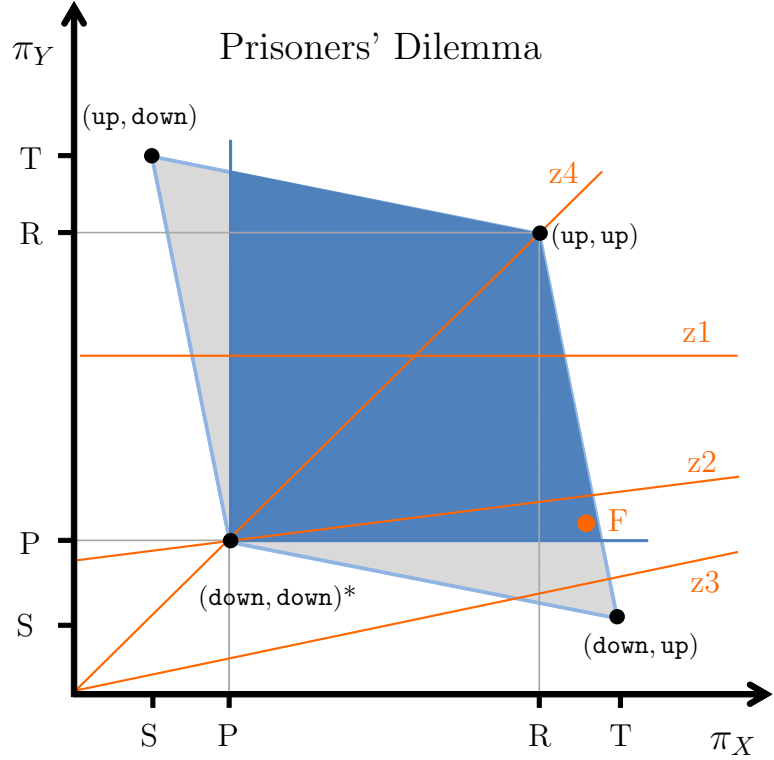
  \caption{Various ZD strategies and the Folk Theorem in iPD}
	\label{fig:ZDvsFolk}
  \floatfoot{Note: The Nash equilibria guaranteed by the Folk Theorem (dark blue) and various ZD constraints (orange). $z1$ represents a \textit{ZDmischief} strategy; $z2$ is a standard \textit{ZDextortion} strategy with $\Delta = P$. As $\chi \rightarrow \infty$, this line will become almost parallel to the $\pi_X$-axis. $z3$ is an \emph{impossible} ZD strategy: for certain parameters the ZD constraints cannot be imposed on the game (since $Y$ does, after all, choose his strategy himself; see \cref{fig:ZDConstraints}). $z4$ is a ``fair'' extortion strategy (such as $\aTFT$) with $\chi=1$. $F$, finally, is the outcome of one particular Nash equilibrium strategy guaranteed by the Folk theorem (one that could be called extortionate).}
\end{figure}

The Folk theorem guarantees the existence of Nash equilibria for a convex set of average payoffs. Therefore, it guarantees Nash equilibria for cooperative payoff distributions as well as for payoffs which could be described as ``unfair,'' such as F in \cref{fig:ZDvsFolk}: provided $\delta$ is close enough to 1, a grim trigger strategy exists to stabilize F as the average payoff of a Nash equilibrium in the infinitely iterated game. Assume $X$ plays this strategy, then $Y$ is confronted with the choice to either comply and receive on average slightly more than $P$, or to play his maximin strategy, receiving $\bar{s}^{(\infty)}_Y = P$.

By contrast, F can also be achieved when $X$ plays a ZD strategy (either mischievous and extortionate), provided $Y$ plays the appropriate strategy in response. In this case, however, F will not be the result of a Nash equilibrium: suppose $X$ plays a \textit{ZDmischief} strategy to yield $\pi_Y=\tilde{\pi}_Y = F_Y$ (where $F_Y$ denotes the Y coordinate of F, i.e\@. $Y$'s expected average payoff if the Nash equilibrium yielding F were played), and $Y$ chooses his strategy such that $\pi_X=F_X$. Then $Y$ will be indifferent to his strategy choice (as discussed earlier), but $X$ will generally be able to increase her payoff (e.g\@. by playing $\aAllD$). Thus, F will not be an equilibrium situation.

Note that $Y$ is free to play a \textit{ZDmischief} strategy himself. Then any combination of two \textit{ZDmischief} strategies is a Nash equilibrium in the infinitely iterated game with $\delta$ large enough, for no player can increase his score by changing his strategy. In an interview, Press notes that this property might be exploited to form cooperative treaties ensuring both parties maximum average payoffs while minimizing incentives to break the treaty \cite{EdgeConv}.

Now assume that $X$ plays a \textit{ZDextortion} strategy, enforcing a linear constraint on which F lies. Then $Y$ can increase his score by moving along the constraint to the outer edge of feasible payoffs (assumed $\delta$ is large enough). Suppose $Y$ does indeed follow these incentives to play $\aAllU$. Then despite $X$'s very successful extortion, the resulting strategy combination is not a Nash equilibrium, for $X$ would again be tempted to play e.g\@. $\aAllD$.

Thus, the outcomes of iterated games with ZD strategies will generally not be Nash equilibria (with the notable exception of two \textit{ZDmischief} strategies), even if the exact same outcomes \emph{could} be achieved via Nash equilibrium strategies guaranteed by the Folk theorem. 

If any desirable outcome achieved by ZD strategies could have been achieved using the Folk theorem, then what is their merit? Aside from the theoretical groundwork they provide, i.e\@. reporting that rather simple \mem{1} strategies can collapse the set of possible scores to arbitrary linear constraints, their primary advantage lies not in stating that a certain strategy combination, if ever achieved, presents an equilibrium situation, but in stating \emph{how} such situations are reached in the long run. This is especially explicit for \textit{ZDextortion} strategies:

Assume $X$ plays a \textit{ZDextortion} strategy with $\chi>1$ and small $\Delta$. Further assume $Y$ is a ``mindless global maximizer'', i.e\@. $Y$ simply tries to maximize his private payoff as best as he can, without trying to change $X$'s behavior. Then if $Y$ knows $X$'s strategy, he will always play his best response, yielding both players (``unfairly'' distributed) maximum payoffs. 

Now assume that $Y$ is less witting: he either does not know $X$'s strategy (i.e\@. he observes her actions, but does not know $\vec{p}$), or he is for some reason unable to immediately find a best response (he cannot solve the complex maximization problem, e.g.). Then $Y$ can be called a ``mindless local maximizer'', i.e\@. he will try to maximize his expected private payoff in a ``trial-and-error''-like fashion by gradually changing parts of his strategy and observing the impacts on his payoff. \PrDy suggest that in many games, a strictly positive gradient leads $Y$ from this situation towards $\aAllU$,%
\footnote{%
          Note that the same holds true for the case of $\aTFT$ (a special ZD strategy).
          }
i.e\@. $\frac{\partial \pi_Y}{\partial \vec{q}} > 0$ for any $\vec{q}$ \cite[pp. 3f.]{PrDy}.%
\footnote{%
          There is an important implication here: assuming $Y$ to maximize either $\pi_Y (\vec{p}, \vec{q})$ or $\hat{\mathscr{P}}_Y (\vec{p}, \vec{q})$, whether locally or globally, implies that $Y$ has full knowledge of $\vec{p}$, or remembers enough outcomes to make an adequate estimate of it. In particular, not all adaptive strategies will be drawn to $\aAllU$: consider the case of learner strategies, such as reinforcement learners or impulse learners \cite{SeltenLearning}. Since such strategies adapt to their observations of how successful certain strategy choices were in the past, they fail to recognize infinitesimal advantages that only matter in the infinite time horizon (the action $a_Y$ in the $t$-th iteration changes the probability distribution over $\mathbb{S}$ in \emph{all} following iterations). Learner strategies will thus generally fail to be lead to $\aAllU$.
}

If this is indeed the case, then instead of stating that F can be the outcome of a Nash equilibrium, it can now be stated that F \emph{will} be achieved if $Y$ adapts his strategy to react to the environment enforced by $X$'s ZD strategy.

Hence, by contrast to the equilibria guaranteed by the Folk theorem, ZD strategies do not present their opponent with an ``all-or-nothing''-choice. Instead, they impose a linear constraint and allow their opponent to move freely along it, hoping to provide enough incentives to nudge him towards favorable behavior. By choosing a ZD strategy, players generally decide \emph{against} a myopically better alternative to maintain the ``carrot-and-stick'' logic leading their opponent towards exploitable behavior. As such, ZD strategies defeat the logic of Nash equilibria and the Folk theorem: they simply have different intentions.

Choosing to play ZD strategies generally appears to be the choice of a very conscious player whose reasoning goes beyond the logic of Nash equilibria: the formal argument that playing a ZD strategy is mostly \emph{not} a best reply would probably not concern a true extortioner who realizes that by changing his strategy he also changes the strategy of his best-response-playing opponent. In a way, ZD strategies are thus exploiting economic rationality, at least the prevalent \textit{ceteris paribus} kind: ZD strategies, in classic economic terms ``irrational,'' exploit best responders. 

Any ZD strategy player $X$ who sticks to her strategy, discarding other ``best responses,'' will likely have a theory of mind \cite[p. 3]{PrDy}. As such, her choice of playing a ZD strategy will depend on whether she attributes her opponent, $Y$, the same meta-cognitive capabilities \cite{PlatoTOM}: if both players have a theory of mind, both might refuse to give in to their opponent's extortion, resulting in a long ultimatum game where both players constantly decline the unfair share offered, resulting in very low scores for both (or, more optimistically put, both players will at some point arrive at a ``fair'' distribution). The success of any ZD strategy thus depends on properties of its opponent that lie beyond the definition of the infinitely iterated game.

	\chapter{Evolutionary Competition}
  \label{chap:EvolComp}

Until now, we merely discussed the existence of ZD strategies, and their implications if $X$ were to play one. We were interested in how the discovery of ZD strategies can ``change the game,'' simply supposing that $X$ be mischievous enough to choose a ZD strategy. What we did not ask yet is, ``\emph{why} would $X$ want to play a ZD strategy?''

Of course, as discussed in \cref{sec:BestResponse,sec:ZDFolk}, if $X$ possesses a superior theory of mind and knows $Y$ to mindlessly maximize his score, \textit{ZDextortion} strategies allow her extraordinarily high payoffs. Another possible area for applications of ZD strategies is evolutionary game theory.

The analyses of classic economics are usually \textit{ceteris paribus}: everything else equal, actors prefer to have more private welfare rather than less. The individual's environment is thus taken as exogenous and fixed: often, however, ``the environment of an individual is itself composed of other individuals who are subject to the same forces of [competition]'' \cite[p. xiii]{Weibull}. Suppose the iterated game $X$ is playing is a competition, e.g\@. for market dominance or for ``employee of the month'' status. Then $X$ will realize that her success in the competition may depend on the peculiarities of her environment at the very time-instant of her move: market success of a firm depends not only on the quality of its products but also on what alternatives are competing for the same market. The possibility of standing out in the workplace depends just as much on how hard the coworkers are trying as on own effort. In economic competition, much like in evolutionary biology, the winner is not always the optimal candidate, but simply the one that outperforms the others.

To enable analysis for this paradigm, the field of evolutionary game theory was developed, both by economists and evolutionary biologists \cite{Weibull}. One implication of modeling competitive situations in selective environments (such as markets or evolution) is that the assumption of well-informed, ``global maximizer'' agents is dropped:%
\footnote{%
          The assumption that agents possess only very limited cognitive abilities and information stems from biologists studying the evolution of lower animals. It may however be quite applicable in some asymmetric economic contexts as well: ``The lower animals are less likely than humans to think about the strategies of their opponents at each stage of a game. Their strategies are more likely to be preprogrammed and their strategy sets more restricted than the businessman's, if perhaps not more so than his customer's'' \cite[p. 143]{RasmusenGT}.
          }
competitors do not know what the optimal response to their environment is (because it constantly changes, for example), and are thus forced to try out different alternatives: the selective forces of evolution then weed out unlucky tries.%
\footnote{%
          It should be noted that \PrDy applied the term ``evolutionary'' to a ``mindless local maximizer'' strategy, i.e\@. to a player who adapts his strategy to optimize his payoff within his environment (``fitness landscape''). While closely related, this use is misleading in the context of evolutionary game theory, where players stick to their strategies, and it instead is evolutionary mutation and selection which leads to what could be called ``local maximization.'' Particularly, the difference is whether the ZD player is herself subject to evolutionary selection. \PrDy's use of ``evolutionary'' is more apt in the context of the iterated game tournaments described in \cref{sec:Tournaments}: \cite{StewartPlotkin} gives a first analysis of ZD strategies in this context.
          }

Replicator dynamics are a standard tool of evolutionary game theory encapsulating these ideas: they state that at any point in time the growth rate of a population share using a certain strategy equals the difference between that strategy's payoff and the complete population's average payoff \cite[p. 73]{Weibull}. The goal for an optimal strategy is thus shifted from finding a global maximum in private welfare to trying to earn more than others to ensure its (genetic, market) survival and dominance. When the global maximum is not known, one cannot be certain that one cannot be outperformed by an opponent. Then it becomes rational to try to make sure that one's opponents earn less than oneself, even if this is only possible at a personal cost.

In this environment, \textit{ZDextortion} strategies with $\chi>1$ are of particular interest: suppose the total population of strategies consists of an incumbent strategy representing a large share of the population and an invader strategy which is only a small part of the population (such an invader may be the result of biological mutation, business innovation or migration in spatial models). Then by making sure that the invader receives smaller payoffs, the incumbent may protect itself against the invader. Conversely, by enforcing higher payoffs for itself, the invader may win an upper hand over an incumbent.

The available literature shows that the evolutionary success of \textit{ZDextortion} strategies is rather limited \cite{AdamiHintze,SigmundNowakExtort}. 
This is due to the fact that \textit{ZDextortion} strategies with $\chi>1$ do not fare well against each other: for illustration, imagine an evolutionary tournament of iPD: players are allotted a fixed strategy and reproduce depending on their payoffs. Suppose an incumbent population playing any strategy INC is invaded by a small group of \textit{ZDextortion} players. Assume the ZD players to have chosen $\Delta=P, \chi=10$: then in playing against each other, they will receive $u_{ZD}(ZD)=\Delta$, where I define $u_x(y)$ as an auxiliary function giving the payoff of strategy $x$ when playing against strategy $y$. By partially giving in to extortion, the incumbent can achieve $u_{\text{INC}}(ZD) \geq P$. Therefore, the extortioner will do worse against itself than the incumbent to be extorted: it is clear that under these circumstances no \textit{ZDextortion} strategy can ever drive an incumbent to extinction \cite{AdamiHintze}. In particular, if the incumbent is a $\aTFT$ strategy, it will achieve cooperative payoffs against itself while $u_{ZD}(ZD)=u_{ZD}(\aTFT)=\Delta$. Thus, a ZD strategy cannot outperform TFT in evolutionary contests. Against some strategies, however, ZD strategies will be able to successfully invade a population and maintain a stable share $\omega<1$ there.%
\footnote{%
          Suppose, for example the incumbent plays $\aAllU$. Then the invading extortioners will successfully outperform the ``naive'' incumbent, growing in population share. As this happens, however, the frequency of ZD strategies playing each other will increase, lowering their average score. At some point, the extortionate population will receive the same average score as the ``naive'' population, yielding a stable population share. To be concise, let $\omega$ be the share of ZD players (then $1-\omega$ is the share of $\aAllU$ players). Then by solving the equations for average score we obtain
          $\omega u_{ZD}(ZD) + (1-\omega) u_{ZD}(\aAllU) = \omega u_{\aAllU}(ZD) + (1-\omega) u_{\aAllU}(\aAllU)$
          \[
          \Rightarrow
          \omega = \frac {u_{ZD}(\aAllU) - u_{\aAllU}(\aAllU)}
                  {u_{ZD}(\aAllU) - u_{\aAllU}(\aAllU))
                   + u_{\aAllU}(ZD) - u_{ZD}(ZD)}
          < 1
          \]
          Note that this does not take into account mutation dynamics, which further contribute to making ZD strategies unstable.
          }

To improve the evolutionary performance of ZD strategies, kin selection mechanisms have been discussed: if the population of \textit{ZDextortion} strategies could identify those players playing the same strategy (i.e\@. realize that their opponent also has a theory of mind), they could play more cooperative strategies against each other. For a more detailed account of the evolutionary performance of ZD strategies, see \cite{SigmundNowakExtort}.

	\chapter{Discussion}
  \label{chap:Discuss}

Some of the popular reception of ZD strategies has been hyperbolic: one (likely influential) example of this is \cite{EdgePoundstone}: ``Press and Dyson have shown that cleverness and unfairness triumph after all [...] [They]'re showing how to fake out evolution!'' As discussed in this paper, the performance of ZD strategies is highly dependent on a number of conditions, and ZD strategies will not generally be successful in unpredictable tournaments; in particular, in evolutionary settings as defined in evolutionary game theory, ZD strategies are not overly successful. By noting that TFT-like strategies generally perform better than other ZD strategies in contexts where strategies are forced to play against themselves as well as others, it is safe to say that extortionate ZD strategies do \emph{not} end the reign of ``cooperative'' strategies like TFT. 

In ``evolutionary'' settings as defined by \PrDy (ZD versus mindless maximizers), strategies that rationally adapt to their environment can successfully be exploited, however: this may be interpreted as a good reason against pure \textit{ceteris paribus} maximization. 

The study of ZD strategies enables exciting new perspectives in the study of iterated 2x2 games. While their existence is not entirely surprising (as suggested by \cref{sec:ZDFolk}), and their performance highly dependent on factors both inside (payoff matrix) and outside the iterated game (theory of mind), ZD strategies highlight important mechanisms in iterated games: the role of memory; classical equilibrium selection and maximizing paths along payoff gradients; the importance of ``uncorrelated asymmetries,'' viz\@. the question who sets his strategy first and who reacts to it; constraints on both players' payoffs, both trivial (such as $\aAllD$) and nontrivial; and many more. Not least of these, the exploitability of ``mindless economic maximizers'' by a player with a mind is an interesting philosophical datum to ponder about---as is the reassuring result that in evolutionary contexts mischief and extortion are limited by their own ``evil.''

\clearpage
\addcontentsline{toc}{chapter}{List of Figures}
\listoffigures
\begingroup  
\let\clearpage\relax
\listoftables
\endgroup


  \appendix
	\appendixpage
  \addappheadtotoc
  \settocdepth{chapter}

	\chapter{Derivation of ZD strategies}
  \label{app:ZDstratDerivation}
  
  \section{ZDmischief strategies}
Feasible parameters for \textit{ZDmischief} strategies are derived for a general symmetric game and a general Battle of the Sexes game.
  
  \subsection{Symmetric games}
For any symmetric game as specified in \ref{tab:symmetricGame}, $\vec{s_Y} = (R, T, S, P)^\T$. Since the goal is to set $\pi_Y=\tilde{\pi}_Y=- \gamma / \beta$, we can use \cref{eq:GeneralZDstrat} with $\gamma = - \tilde{\pi}_Y \beta$. This yields the general \textit{ZDmischief} strategy%
  
  \[
    \vec{p} = 
      \begin{pmatrix}
        \beta (R-\tilde{\pi}_Y) + 1 \\
        \beta (T-\tilde{\pi}_Y) + 1 \\
        \beta (S-\tilde{\pi}_Y)     \\
        \beta (P-\tilde{\pi}_Y)     \\
      \end{pmatrix}	
  \]
For symmetric games where $\min \{R, T\} \geq \max \{S, P\}$, $\vec{p}$ is thus in the realm of probability vectors for %
  \[
    \max \{S, P\} \leq \tilde{\pi}_Y \leq \min \{R, T\}
  \]
 \qed

  \subsection{General Battle of the Sexes}
For any BS game as specified in \ref{tab:BoSGame}, $\vec{s_Y} = (D, C, L, F)^\T$. Since the goal is to set $\pi_Y=\tilde{\pi}_Y=- \gamma / \beta$, we can use \cref{eq:GeneralZDstrat} with $\gamma = - \tilde{\pi}_Y \beta$. This yields the general \textit{ZDmischief} strategy%
  
  \[
    \vec{p} = 
      \begin{pmatrix}
        \beta (D-\tilde{\pi}_Y) + 1 \\
        \beta (C-\tilde{\pi}_Y) + 1 \\
        \beta (L-\tilde{\pi}_Y)     \\
        \beta (F-\tilde{\pi}_Y)     \\
      \end{pmatrix}	
  \]
It follows that for $\vec{p}$ to be in the realm of probability vectors, %
  \[
    \max \{C, D\} \leq \tilde{\pi}_Y \leq \min \{L, F\} 
    \ \vee \ 
    \max \{F, L\} \leq \tilde{\pi}_Y \leq \min \{C, D\}
  \]
Since in BS as defined here, $L \leq C < D < F$, this is impossible.
  \qed

  \section{ZDextortion strategies}
Feasible parameters for \textit{ZDextortion} strategies are derived for a general symmetric game and a general Battle of the Sexes game.

  \subsection{Symmetric games}

For any symmetric game as specified in \ref{tab:symmetricGame}, $\vec{s_X} = (R, S, T, P)^\T, \vec{s_Y} = (R, T, S, P)^\T$. The goal is to set $(\pi_X - \Delta) = \chi(\pi_Y - \Delta)$, which could be done using \cref{eq:GeneralZDstrat} with the $\gamma$ parameter described in \cref{sec:MischiefZD}. However, derivation is much clearer using the equivalent linear combination from \cite[p. 3]{PrDy}: 
$\vec{p} = \phi [ (\vec{s_X} - \Delta \vec{1}) - \chi (\vec{s_Y} - \Delta \vec{1}) ] + (1, 1, 0, 0)^T$
with normalizing factor $\phi$. This yields the general \textit{ZDextortion} strategy%
  
  \[
    \vec{p} = \phi
      \begin{pmatrix}
        (\chi - 1) (\Delta - R)          \\
        \chi (\Delta - T) - (\Delta - S) \\
        \chi (\Delta - S) - (\Delta - T) \\
        (\chi - 1) (\Delta - P)          \\
      \end{pmatrix}	
      +
      \begin{pmatrix}
        1 \\
        1 \\
        0 \\
        0 \\
      \end{pmatrix}	      
  \]
It is plain that for $\chi = 1$ this is always a feasible strategy. Since 
$\max \{S, P\} \leq \min \{R, T\}$,
no feasible solutions exist for $\chi < 1$, and $\phi$ be positive for $\chi \geq 1$. From this and the equations for $p_1$ and $p_4$ follows the first constraint, $P \leq \Delta \leq R$.

The equations for $p_2$ and $p_3$ further imply that 
$\chi (\Delta - T) \leq (\Delta - S)$ 
and 
$\chi (\Delta - S) \geq (\Delta - T)$. 
This yields%

\[
  \chi \leq 
       \begin{cases}
        \frac{\Delta - S}{\Delta - T}& \Delta > T\\
        \frac{\Delta - T}{\Delta - S}& \Delta < S\\
       \end{cases}
  \ \ \text{, and} \ \ 
  \chi \geq 
       \begin{cases}
        \frac{\Delta - S}{\Delta - T}& \Delta < T\\
        \frac{\Delta - T}{\Delta - S}& \Delta > S\\
       \end{cases}  
\]
Thus, for $S=T$, $\chi>1$ is not feasible. For $S < \Delta < T$, there is no upper bound on $\chi$. For $\Delta > T$ and $\Delta < S$, positive upper bounds on $\chi$ exist.
\qed

Assume further that $\Delta=0$. Then the equations for $p_1$ and $p_4$, and $R>P$ imply that $P=0$. $\chi$ will then be confined to the range $[1, \frac{T}{S}]$.
\qed

  \subsection{General Battle of the Sexes}

For any BS game as specified in \ref{tab:BoSGame}, $\vec{s_X} = (F, C, L, D)^\T, \vec{s_Y} = (D, C, L, F)^\T$. Analogously to the derivation for symmetric games above, we set
$\vec{p} = \phi [ (\vec{s_X} - \Delta \vec{1}) - \chi (\vec{s_Y} - \Delta \vec{1}) ] + (1, 1, 0, 0)^T$. This yields the general \textit{ZDextortion} strategy%
  
  \[
    \vec{p} = \phi
      \begin{pmatrix}
        \chi (\Delta - D) - (\Delta - F) \\
        (\chi - 1) (\Delta - C)          \\
        (\chi - 1) (\Delta - L)          \\
        \chi (\Delta - F) - (\Delta - D) \\
      \end{pmatrix}	
      +
      \begin{pmatrix}
        1 \\
        1 \\
        0 \\
        0 \\
      \end{pmatrix}	      
  \]
It is clear that for $\chi = 1$ this is always a feasible strategy. Since 
$\max \{L, C\} \leq \min \{D, F\}$,
$\phi$ will be negative for $\chi > 0$. From this and the equations for $p_2$ and $p_3$ we derive the first constraint, $C = \Delta = L$. Then from the equations for $p_1$ and $p_4$ we can further follow%

\[
  \frac{\Delta-D}{\Delta-F}
  \leq \chi \leq 
  \frac{\Delta-F}{\Delta-D}
\]
Thus, for any BS game as defined in this paper, either $C=L$ and $\chi$ has both positive upper and lower bounds (this romantically implies that $\chi<1$ is possible in iBS), or $C>L$ and $\chi=1$. 
\qed

	\chapter{Simulation codes in Java}

Various parts of this paper depend on computer simulations. The following sections include the Java codes used to obtain the results. They were compiled on Java 1.7 on Windows 7, but should be fairly compatible.

  \section{Markov chain will converge}
  \label{app:MatrixMult}

The following code simulates the condition given in \cite[§2.2.2]{StroockMarkov} to prove that

\[
\mat{A}_4 := 1/4 \sum_{m=0}^3{\mat{M}^m}
\]
will have at least one column with non-zero entries, where $\mat{M}$ is the Markov transition matrix:
\[
	\mat{M}=
	\begin{bmatrix}
		p_1q_1 & p_1(1-q_1) & (1-p_1)q_1 & (1-p_1)(1-q_1)\\
		p_2q_3 & p_2(1-q_3) & (1-p_2)q_3 & (1-p_2)(1-q_3)\\
		p_3q_2 & p_3(1-q_2) & (1-p_3)q_2 & (1-p_3)(1-q_2)\\
		p_4q_4 & p_4(1-q_4) & (1-p_4)q_4 & (1-p_4)(1-q_4)
	\end{bmatrix}	
\]
This suffices to prove that the Markov process given by $\mat{M}$ converges: $\mat{A}_4$ gives the average number of visits to each state in four subsequent iterations. Recalling that the number of states $\# \mathbb{S}=4$, it is clear that it is possible for the Markov chain to reach any state within four iterations, if the state is to be reached at all. The code finds some cases where $\mat{A}_4$ still has zero elements in every column; inspection shows that such chains \emph{never} visit all states: in this case, the chain may be analyzed using a reduced state set $\mathbb{S'} \subset \{\aClockWise\}$, leaving out all states which are never visited, to obtain a transition matrix for which $\mat{A}_4$ has at least one column with no zero elements.

For illustration, take the case of $\vec{p}=\vec{q}=(1,0,1,0)^\T$, an infinitely iterated game of $\aTFT$ vs. $\aTFT$. The stationary distribution $\vec{\pi}$ then depends entirely on the starting distribution $\vec{\mu_1}$: if the game's first iteration turns out to be $(\aUP, \aUP)$ or $(\aDOWN, \aDOWN)$, the game will forever stay in this state, yielding $\vec{\pi} = (1, 0, 0, 0)$ or $\vec{\pi} = (0, 0, 0, 1)$, respectively. The chain is then reduced to a single state. If the game's first iteration is $(\aUP, \aDOWN)$ or $(\aDOWN, \aUP)$, the game will forever alternate between those two states, yielding $\vec{\pi} = (0, 0.5, 0.5, 0)$ and reducing the chain to two states. Different values for $\vec{\mu_1}$ will yield linear combinations of those three vectors.

\begin{tiny}
\begin{lstlisting}[language=Java,caption=MatrixMult.java]
/**
 * 	Simulation of a Markov chain as presented in Press and Dyson 2012.
 *  We show that the chain will converge using sect. 2.2.2 of Stroock 2005, by
 *  showing that A := 1/4 SUM_{m=0}^{3} {M^m} will always have at least one
 *  column with non-zero entries.
 */
package de.LarsRoemheld.MatrixMult;

/**
 * @author Lars Roemheld
 *
 */
public class MatrixMultiplication {

	/*
	 * Multiply two matrices, given as arrays
	 */
	static double[][] MatrixMultiply(double a[][], double b[][]) {
		int aRows = a.length,
				aColumns = a[0].length,
				bRows = b.length,
				bColumns = b[0].length;

		if ( aColumns != bRows ) {
			throw new IllegalArgumentException("A:Rows did not match B:Columns.");
		}

		double[][] result = new double[aRows][bColumns];

		for(int i = 0; i < aRows; i++) { // aRow
			for(int j = 0; j < bColumns; j++) { // bColumn
				for(int k = 0; k < aColumns; k++) { // aColumn
					result[i][j] += a[i][k] * b[k][j];
				}
			} 
		}

		return result;
	}

	static double[][] MatrixAdd(double[][] m1, double[][] m2) {

		if ( m1.length != m2.length ) {
			throw new IllegalArgumentException("MatrixAdd: Different matrix lengths.");
		}

		double[][] result = new double[m1.length][m1[0].length];

		for (int row=0; row<m1.length; row++) {
			for (int col=0; col<m1[row].length; col++) {
				result[row][col] = m1[row][col] + m2[row][col];
			}
		}

		return result;
	}


	
	public static void main(String[] args) {
		final int MAX_Power = 5;

		System.out.println("started calculations. . .");
		System.out.println("Strategy combination that do not converge regardless of starting point:");
		
		double M[][] = new double[5][5];
		double R[][] = new double[5][5];		
		double A[][] = new double[5][5];		

		double E[][] = new double[5][5]; // E_4 unity matrix
		E[1][1] = 1;
		E[2][2] = 1;
		E[3][3] = 1;
		E[4][4] = 1;

		int maxM = -1;
		
		// Loop through all relevant p, q combinations: 1s and 0s
		for (double p1 = 0; p1 <= 1; p1+=1) {
		for (double p2 = 0; p2 <= 1; p2+=1) {
		for (double p3 = 0; p3 <= 1; p3+=1) {
		for (double p4 = 0; p4 <= 1; p4+=1) {

		for (double q1 = 0; q1 <= 1; q1+=1) {
		for (double q2 = 0; q2 <= 1; q2+=1) {
		for (double q3 = 0; q3 <= 1; q3+=1) {
		for (double q4 = 0; q4 <= 1; q4+=1) {			
			// Create Markov transition matrix for p, q
			// As specified in Press and Dyson 2012
			M[1][1] = p1*q1;
			M[2][1] = p2*q3;
			M[3][1] = p3*q2;
			M[4][1] = p4*q4;

			M[1][2] = p1*(1-q1);
			M[2][2] = p2*(1-q3);
			M[3][2] = p3*(1-q2);
			M[4][2] = p4*(1-q4);

			M[1][3] = (1-p1)*q1;
			M[2][3] = (1-p2)*q3;
			M[3][3] = (1-p3)*q2;
			M[4][3] = (1-p4)*q4;

			M[1][4] = (1-p1)*(1-q1);
			M[2][4] = (1-p2)*(1-q3);
			M[3][4] = (1-p3)*(1-q2);
			M[4][4] = (1-p4)*(1-q4);

			// A = E_4
			A = E.clone();
			
			// Check minimum power until A has a column stictly greater than 0
			R = E.clone(); // R = E_4
			
			boolean foundNonZeroColumn = false;
			
			powers_m:
			for (int power_m = 1; power_m < MAX_Power; power_m++) {

				R = MatrixMultiply(M, R);
								
				A = MatrixAdd(A, R);

				for (int j = 1; j<=4; j++) {
					double colmin = Math.min(A[1][j], Math.min(A[2][j], Math.min(A[3][j], A[4][j])));
					
					if (colmin > 0) {
						maxM = Math.max(maxM, power_m);
						foundNonZeroColumn = true;
						break powers_m;
					}

				}

			}

			if (!foundNonZeroColumn)
			{
				System.out.print("p = (" + p1 + ", " + p2 + ", " + p3 + ", " + p4 + ")  --  ");
				System.out.println("q = (" + q1 + ", " + q2 + ", " + q3 + ", " + q4 + ")");
			}
		}
		}
		}
		}

		}
		}
		}
		}
		
		if (maxM == MAX_Power) 
			System.out.println("Please increase MAX_Power.");
		else
		{
			System.out.println("ALL other strategy combinations will converge at least over " + maxM + " iterations.");
		}
	}
}
\end{lstlisting}
\end{tiny}

  \section{Simulation of iterated 2x2 games}
  \label{app:ZDsimulation}

The following classes provide a framework for simulating general iterated 2x2 games. They were used for all numerical simulations in this paper.

\begin{tiny}
\begin{lstlisting}[language=Java,caption=IteratedGame.java]
package de.LarsRoemheld.ItGame;

import java.io.*;

/**
 * This project provides a framework to simulate iterated 2x2 games.
 * 
 * This class uses general implementations (such as a "2x2 stage game" and
 * an "2x2 agent". It simply calls the functionality of these classes and
 * logs the game outcome.
 * 
 * @author Lars Roemheld
 */
public class IteratedGame {
	/* Define payoffs for different games. Payoffs are arrays which give
	 * payoffs for one specific agent in the order 
	 * agent1/agent2: up/up, up/down, down/up, down/down
	 */
	public static final double PAYOFFS_PD_A1[] = {3, 0, 5, 1};
	public static final double PAYOFFS_PD_A2[] = {3, 5, 0, 1};
	private static final String[] OUTCOMENAMES = {"up/up", "up/down", "down/up", "down/down"};
	
	// Provide a path to a log file for file output -- else console output is used.
	private static final String LogFilePrefix = "";// "D:\\iPD_";
	// Number of iterations to simulate
	private static final int ITERATIONS = 10000; 
	// Number of individual games of ITERATIONS iterations to simulate
	private static final int NGAMES = 1;     
	
	private static Agent_2x2 agent1;
	private static Agent_2x2 agent2;
	private static StageGame_2x2 stageGamePlayed;

	
	public static void main(String[] args) {
		// Log to file or simply show results on screen?
		PrintStream log;
		if (LogFilePrefix != "")
		{
			try {
				log = new PrintStream(LogFilePrefix + NGAMES + " times " + ITERATIONS + ".log");
			}
			catch (IOException e) {
				log = System.out;
				log.println("Could not open file");
			}
		}
		else
			log = System.out;
		
		log.println("Averaging simulation of " + NGAMES + " games over " + ITERATIONS + " iterations.");
		log.println("iteration  agent1  agent2");
		System.out.println("Starting simulation of " + NGAMES + " games over " + ITERATIONS + " iterations.");

		double[] a1Scores = new double[ITERATIONS + 1];
		double[] a2Scores = new double[ITERATIONS + 1];
		
		// Simulate a number of games
		for (int game = 1; game <= NGAMES; game++)
		{
		// With new agents and new game each time
		agent1 = new ProbAgent(0, 0.64d, 0.18d, 0.28d, 0.0d, 0.5d);
		agent2 = new ReinforcementLearnerAgent(0, 1d, PAYOFFS_PD_A1); // Payoffs for A1 since outcomes always seen from his perspective!
		
		stageGamePlayed = new StageGame_2x2(PAYOFFS_PD_A1, PAYOFFS_PD_A2);
		
		// Simulate all iterations and store average scores.
		for (int it = 1; it <= ITERATIONS; it++) {
			int outcome = stageGamePlayed.simGame(agent1, agent2);
			
			a1Scores[it] = (a1Scores[it] * (game - 1) + agent1.getFitness() / it) / game;
			a2Scores[it] = (a2Scores[it] * (game - 1) + agent2.getFitness() / it) / game;
			// if (it % 10000 == 0) log.println("Learner-pUP: " + ((ReinforcementLearnerAgent) agent2).getpUP());
			// log.println("Game result: " + OUTCOMENAMES[outcome-1] + ". Average scores after " + it + " iterations: agent 1: " + agent1.getFitness() / it + " -- agent 2: " + agent2.getFitness() / it);
		}
		}
		
		System.out.println("Starting output.");
		for (int it = 1; it <= ITERATIONS; it++) 
		{
			log.println(it + "  " + a1Scores[it] + "  " + a2Scores[it]);
		}
		
		System.out.println("done.");
	}
}
\end{lstlisting}
\begin{lstlisting}[language=Java,caption=StageGame\_2x2.java]
package de.LarsRoemheld.ItGame;

/**
 * Provides a single one-shot 2x2 game to enable simulation of iterated games.
 * @author Lars Roemheld
 */
public class StageGame_2x2 {
	private double gamePAYOFFS_a1[];
	private double gamePAYOFFS_a2[];
	private Moves move1;
	private Moves move2;
	private int gameOutcome;
	
	public enum Moves {
		UP, DOWN
	}
	
	/**
	 * Constructor.
	 * @param payoffs_a1 agent1's payoffs in this game. An array containing the game payoffs as [agent1/agent2: up/up, up/down, down/up, down/down]
	 * @param payoffs_a2 agent2's payoffs in this game. An array containing the game payoffs as [agent1/agent2: up/up, up/down, down/up, down/down]
	 */
	public StageGame_2x2(double payoffs_a1[], double payoffs_a2[]) {
		this.gamePAYOFFS_a1 = (double[]) payoffs_a1.clone();
		this.gamePAYOFFS_a2 = (double[]) payoffs_a2.clone();
	}
	
	/**
	 * Simulates a single game between two agents.
	 * @return Returns the game outcome as integer 1..4, as indices of [up/up, up/down, down/up, down/down]
	 */
	public int simGame(Agent_2x2 a1, Agent_2x2 a2) {
		move1 = a1.move(); // Make the agents move
		move2 = a2.move();
		
		gameOutcome = 0; // Interpret the results
		if (move1 == Moves.UP && move2 == Moves.UP) 
			{ gameOutcome = 1;}  // uu
		else if (move1 == Moves.UP && move2 == Moves.DOWN)
			{ gameOutcome = 2;}  // ud
		else if (move1 == Moves.DOWN && move2 == Moves.UP)
			{ gameOutcome = 3;}  // du
		else if (move1 == Moves.DOWN && move2 == Moves.DOWN)
			{ gameOutcome = 4;}  // dd

		// Pay out payoffs
		a1.addFitness(gamePAYOFFS_a1[gameOutcome - 1]);
		a2.addFitness(gamePAYOFFS_a2[gameOutcome - 1]);
		
		// Notify agents of the game outcome
		a1.outcome(move1, move2);
		a2.outcome(move2, move1);
		
		return gameOutcome;
	}
}
\end{lstlisting}
\begin{lstlisting}[language=Java,caption=Agent\_2x2.java]
package de.LarsRoemheld.ItGame;

/*
 * This provides an abstract class for agents in iterated 2x2 games:
 * Any agent in 2x2 stage games should implement this class.
 */
public abstract class Agent_2x2 {
	// The private scoreboard
	private double fitness;
	
	// Set up a new agent and endow him with initial fitness.
	protected Agent_2x2(double fitness) {
		this.fitness = fitness;
	}
	
	public double getFitness() {
		return this.fitness;
	}
	public void setFitness(double fitness) {
		this.fitness = fitness;
	}	
	public void addFitness(double dFitness) {
		this.fitness = this.fitness + dFitness;
	}
	
	public abstract StageGame_2x2.Moves move();
	public abstract void outcome(StageGame_2x2.Moves myMove, StageGame_2x2.Moves opponentMove);
}
\end{lstlisting}
\begin{lstlisting}[language=Java,caption=ProbAgent.java]
package de.LarsRoemheld.ItGame;

/*
 * This class implements a memory-1 probabilistic strategy for 
 * iterated 2x2 games.
 */
public class ProbAgent extends Agent_2x2{
	// This agent's strategy is given by four conditional. . .
	protected double p1;
	protected double p2;
	protected double p3;
	protected double p4;
	// and one unconditional probability (see constructor)
	protected double pNoMem; 
	
	/* This agent's memory: the last game's outcome as seen 
	 * by him (i.e. assuming he is agent1)                  */
	private int memory_one;
	
	/**
	 * Set up a new probabilistic agent
	 * @param fitness starting fitness
	 * @param p1 prob(up | (up/up) played before)
	 * @param p2 prob(up | (up/down) played before)
	 * @param p3 prob(up | (down/up) played before)
	 * @param p4 prob(up | (down/down) played before)
	 * @param pNoMem prob(up | no memory of last move, e.g. in first move)
	 */
	public ProbAgent(double fitness, double p1, double p2, double p3, double p4, double pNoMem) {
		super(fitness);
		
		this.p1 = p1;
		this.p2 = p2;
		this.p3 = p3;
		this.p4 = p4;
		this.pNoMem = pNoMem;
		
		this.memory_one = -1;
	}
	
	public StageGame_2x2.Moves move() {
		// Determine the probability of playing UP
		double probU;
		switch (this.memory_one) {
			case 1: probU = this.p1; break;
			case 2: probU = this.p2; break;
			case 3: probU = this.p3; break;
			case 4: probU = this.p4; break;
			default: probU = pNoMem; break;	
		}
		
		// Roll the dice (a random number between 0 and 1)
		double fate = Math.random();
		
		return (fate < probU) ? StageGame_2x2.Moves.UP : StageGame_2x2.Moves.DOWN;
	}
	
	/**
	 * Receive the last game outcome (to memorize)
	 * @param myMove last move by this agent
	 * @param opponentMove last move by this agent's last opponent
	 */
	public void outcome(StageGame_2x2.Moves myMove, StageGame_2x2.Moves opponentMove) {
		this.memory_one = -1;
		
		if (myMove == StageGame_2x2.Moves.UP && opponentMove == StageGame_2x2.Moves.UP) 
			{ this.memory_one = 1;}  // uu
		else if (myMove == StageGame_2x2.Moves.UP && opponentMove == StageGame_2x2.Moves.DOWN)
			{ this.memory_one = 2;}  // ud
		else if (myMove == StageGame_2x2.Moves.DOWN && opponentMove == StageGame_2x2.Moves.UP)
			{ this.memory_one = 3;}  // du
		else if (myMove == StageGame_2x2.Moves.DOWN && opponentMove == StageGame_2x2.Moves.DOWN)
			{ this.memory_one = 4;}  // dd
	}
}
\end{lstlisting}
\begin{lstlisting}[language=Java,caption=ReinforcementLearnerAgent.java]
package de.LarsRoemheld.ItGame;

/**
 * Implements a 2x2-game agent utilizing a simple learning 
 * algorithm (reinforcement learning) based on Chmura, Goerg, Selten 2012
 * (playing memory-0 there).
 * 
 * @author Lars Roemheld
*/
public class ReinforcementLearnerAgent extends ProbAgent {
	protected double pUP;
	
	// A memory of sorts: total scores gained when playing UP or DOWN
	protected double scoresUP;   
	protected double scoresDOWN;
	protected StageGame_2x2.Moves myLastMove = null;
	protected double myDiscountFactor;
	
	protected double[] game_payoffs;
	
	public double getpUP() {
		return pUP;
	}

	/**
	 * Create a new learner agent with
	 * @param fitness Starting fitness
	 * @param discountFactor A discount factor determining how much of the 
	 * 				following iteration's payoffs are attributed to this iteration's action
	 * @param payoffs The game's payoffs as seen from this agent's perspective
	 */
	public ReinforcementLearnerAgent(double fitness, double discountFactor, double[] payoffs) {
		super(fitness, 0.5, 0.5, 0.5, 0.5, 0.5);
		
		this.pUP = 0.5d;
		this.scoresDOWN = 0d;  this.scoresUP = 0d;
		this.myDiscountFactor = discountFactor;
		this.game_payoffs = payoffs.clone();
	}

	// Record success of different actions
	@Override
	public void outcome(StageGame_2x2.Moves myMove, StageGame_2x2.Moves opponentMove) {
		double payoff = 0;
		
		if (myMove == StageGame_2x2.Moves.UP) {
			if (opponentMove == StageGame_2x2.Moves.UP) payoff = this.game_payoffs[0]; // uu
			else payoff = this.game_payoffs[1]; } // ud
		else {
			if (opponentMove == StageGame_2x2.Moves.UP) payoff = this.game_payoffs[2]; // du
			else payoff = this.game_payoffs[3]; } // dd		
		
		// Record success of this move
		if (myMove == StageGame_2x2.Moves.UP)   this.scoresUP += payoff;
		if (myMove == StageGame_2x2.Moves.DOWN) this.scoresDOWN += payoff;
		
		// And take into account this result for attractiveness 
		// of last move, weighted by discountFactor
		if (myLastMove == StageGame_2x2.Moves.UP)   this.scoresUP += this.myDiscountFactor * payoff;
		if (myLastMove == StageGame_2x2.Moves.DOWN) this.scoresDOWN += this.myDiscountFactor * payoff;
			
		// If we have gathered enough experience with both actions, start "learning"
		if (this.scoresUP > 0 && this.scoresDOWN > 0)
			this.pUP = this.scoresUP / (this.scoresUP + this.scoresDOWN);
		
		this.p1 = this.pUP; this.p2 = this.pUP; this.p3 = this.pUP; this.p4 = this.pUP;
		this.pNoMem = this.pUP;
		
		myLastMove = myMove;
	}
}
\end{lstlisting}
\end{tiny}

\end{document}